\documentclass[
aps,
physrev,
superscriptaddress,
amsmath,
amssymb,
nofootinbib,
floatfix,
twocolumn,
reprint
]
{revtex4-2}

\usepackage[utf8]{inputenc}
\usepackage{hyperref}
\usepackage{graphicx,color}
\usepackage{url}
\usepackage{bm}
\usepackage[italicdiff]{physics}
\usepackage[all]{xy}
\usepackage{qcircuit}
\usepackage{amsthm}
\usepackage{enumitem}
\usepackage{comment}
\setlist[description]{leftmargin=\parindent,labelindent=\parindent}

\theoremstyle{plain}

\newtheorem*{thm*}{Theorem}
\newtheorem*{prop*}{Proposition}

\usepackage{hyperref}
\hypersetup{
colorlinks=true, 
linkcolor=blue, 
citecolor=blue 
}
  
\begin{document}
 
\title{Quantum many-body simulation of finite-temperature systems with sampling\\ a series expansion of a quantum imaginary-time evolution}

\author{Norifumi Matsumoto}
\affiliation{Quantum Laboratory, Fujitsu Research, Fujitsu Limited., 4-1-1 Kamikodanaka, Nakahara, Kawasaki, Kanagawa 211-8588, Japan}
\email{m.norifumi@fujitsu.com}

\author{Shoichiro Tsutsui}
\affiliation{QunaSys Inc., Aqua Hakusan Building 9F, 1-13-7 Hakusan, Bunkyo, Tokyo 113-0001, Japan}

\author{Yuya O. Nakagawa}
\affiliation{QunaSys Inc., Aqua Hakusan Building 9F, 1-13-7 Hakusan, Bunkyo, Tokyo 113-0001, Japan}

\author{Yuichiro Hidaka}
\affiliation{QunaSys Inc., Aqua Hakusan Building 9F, 1-13-7 Hakusan, Bunkyo, Tokyo 113-0001, Japan}

\author{Shota Kanasugi}
\affiliation{Quantum Laboratory, Fujitsu Research, Fujitsu Limited., 4-1-1 Kamikodanaka, Nakahara, Kawasaki, Kanagawa 211-8588, Japan}

\author{Kazunori Maruyama}
\affiliation{Quantum Laboratory, Fujitsu Research, Fujitsu Limited., 4-1-1 Kamikodanaka, Nakahara, Kawasaki, Kanagawa 211-8588, Japan}

\author{Hirotaka Oshima}
\affiliation{Quantum Laboratory, Fujitsu Research, Fujitsu Limited., 4-1-1 Kamikodanaka, Nakahara, Kawasaki, Kanagawa 211-8588, Japan}

\author{Shintaro Sato}
\affiliation{Quantum Laboratory, Fujitsu Research, Fujitsu Limited., 4-1-1 Kamikodanaka, Nakahara, Kawasaki, Kanagawa 211-8588, Japan}

\date{\today}

\begin{abstract}

Simulating thermal-equilibrium properties at finite temperature is crucial for studying quantum many-body systems. Quantum computers are expected to enable us to simulate large systems at finite temperatures, overcoming challenges faced by classical computers, \textcolor{black}{such as} the sign problem of the quantum Monte-Carlo technique. Conventional methods suitable for fault-tolerant quantum computing (FTQC) devices are designed for studying large-scale quantum many-body systems but require a large number of ancilla qubits and a deep quantum circuit with many basic gates, making them unsuitable for the early stage of the FTQC era, at which the availability of qubits and quantum gates is limited. In this paper, we propose a method suitable for quantum devices in this early stage to calculate the thermal-equilibrium expectation value of an observable at finite temperatures.
Our proposal, named the Markov-chain Monte Carlo with sampled pairs of unitaries (MCMC-SPU) algorithm, involves sampling simple quantum circuits and generating the corresponding statistical ensembles. This approach addresses the issues of resource demand and the decay in probability associated with postselection of measurement outcomes on ancilla qubits.
We validate our proposal with numerical simulation on the one-dimensional transverse-field Ising model as an illustrative example.

\end{abstract}

\maketitle

\begin{figure*}[htbp]
    \includegraphics[keepaspectratio, scale=0.54]{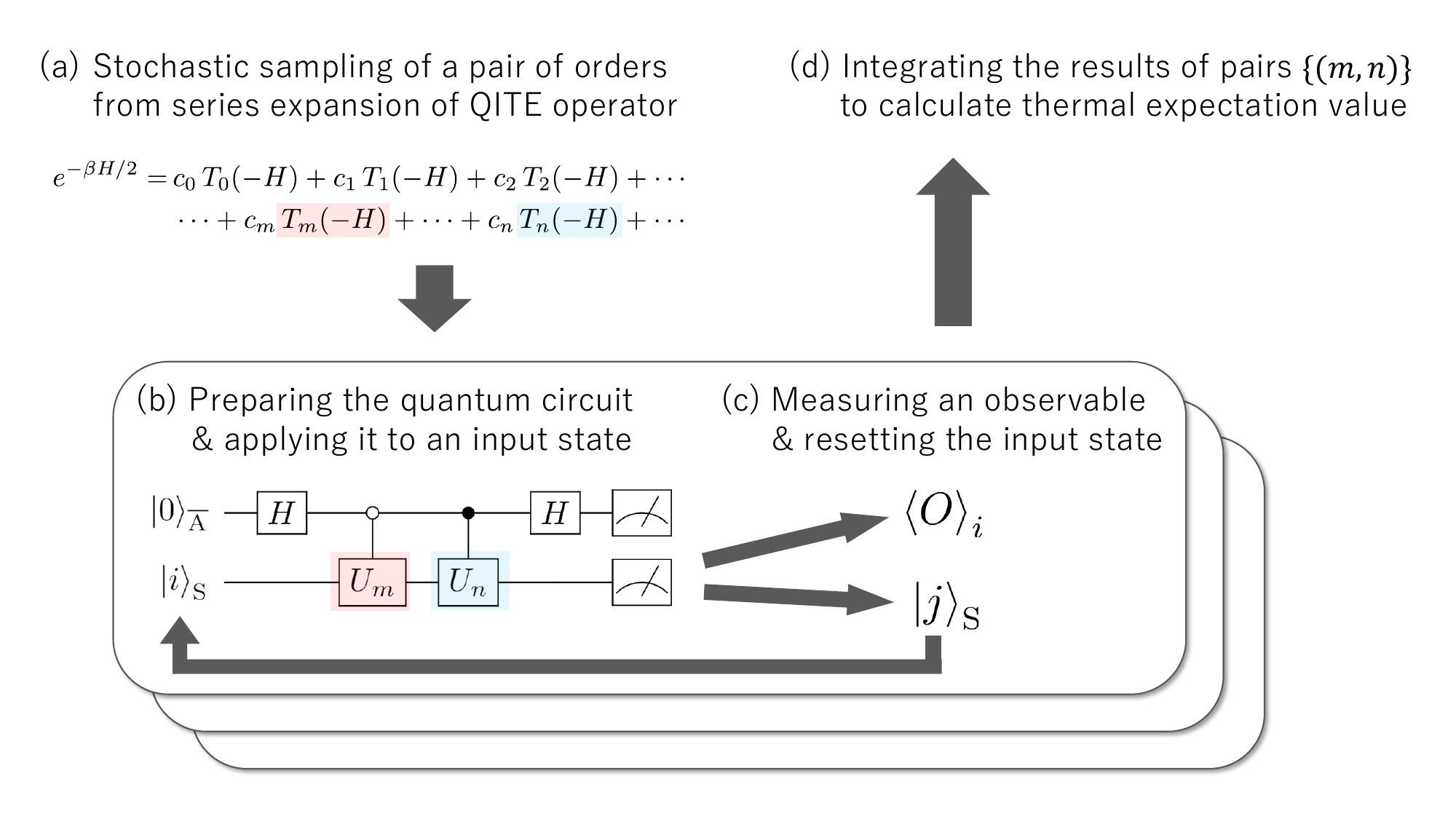}
    \caption{
    Overview of our proposal of the Markov-chain Monte Carlo with sampled pairs of unitaries (MCMC-SPU) algorithm.
    (a) Our proposal is based on a series expansion of a quantum imaginary-time evolution (QITE) operator. We perform stochastic sampling of a pair of orders $m$ and $n$ of polynomials from the series expansion.
    (b) For each pair of orders stochastically sampled in (a), we prepare a corresponding quantum circuit and apply it to a randomly chosen input state in the computational basis.
    (c) We measure an observable with respect to the state obtained in (b), and then reset the input state by performing a projective measurement on the output state of (b), the result of which is used as the input state for the next cycle.
    (d) We finally integrate the results of a large number of stochastic samples of pairs of orders $m$ and $n$ to calculate the thermal-equilibrium expectation value of the observable at finite temperature.
    }
    \label{fig:overview-MCMCSPU}
\end{figure*}

\section{Introduction}\label{sec:Intro}

Recent advances in quantum control and quantum computing~\cite{Ekert1996, Steane1998, Ladd2010} have attracted widespread attention~\cite{Cao2019, Gyongyosi2019, Orus2019, Outeiral2021, Gill2022}. 
One of the most promising applications of quantum computers is the simulation of quantum many-body systems~\cite{Lesenstein2007, Buluta2009, Kassal2011, Georgescu2014, OMalley2016, Zohar2016, Noh2017, Hempel2018, Cao2019, Kokail2019, Bauer2020, Bauer2023}, which plays a crucial role in various fields of science such as condensed-matter physics~\cite{Lesenstein2007, Buluta2009, Georgescu2014, Noh2017}, quantum chemistry~\cite{Buluta2009, Kassal2011, Georgescu2014, OMalley2016, Hempel2018, Cao2019, Bauer2020}, and high-energy physics~\cite{Buluta2009, Gerritsma2010, Georgescu2014, Zohar2016, Kokail2019, Bauer2023}. A key task in quantum simulation is the evaluation of the ground-state energy of these systems. Techniques such as the variational quantum eigensolver (VQE)~\cite{Peruzzo2014, McClean2016, Kandala2017, Moll2018} enable us to estimate the ground-state energy using near-term noisy intermediate-scale quantum (NISQ) devices~\cite{Preskill2018, Cerezo2021, Bharti2022}.
Furthermore, quantum phase estimation (QPE)~\cite{Kitaev1995, Cleve1998, Abrams1999, Higgins2007, Dorner2009} offers a more precise estimation of the ground-state energy on fault-tolerant quantum computing (FTQC) devices.

Another important task is simulating the properties of quantum many-body systems at finite temperatures.
Classical computation techniques, such as the quantum Monte-Carlo method~\cite{Ceperley1986, Sandvik1991, Foulkes2001, Austin2012, Carlson2015}, the minimally-entangled typical thermal state (METTS) algorithm~\cite{White2009, Stoudenmire2010}, and the thermal pure quantum (TPQ) states~\cite{Sugiura2012, Sugiura2013, Hyuga2014, Endo2018, Iwaki2021, Tsutsui2022}, have been developed for this purpose.
Recently, methods to implement finite-temperature simulation techniques on quantum devices~\cite{Motta2020, Lu2021, Powers2022, Coopmans2023, Schuckert2023, Summer2024} have been proposed. These implementations are expected to overcome the challenges of simulating large systems at finite temperatures, where quantum Monte-Carlo methods on classical computers face the sign problem~\cite{Loh1990, Troyer2005, Alexandru2022}. Furthermore, some of the proposals have been experimentally realized for small-sized quantum devices~\cite{Motta2020, Sun2021, Bassman2021, Yeter2021, Powers2022, Hemery2023, Summer2024}.

In the simulation of quantum many-body systems at finite temperatures on quantum devices, quantum imaginary-time evolution (QITE) plays a crucial role in realizing the canonical statistical ensemble, also known as the Gibbs state (see Sec.~\ref{sec:Review} for details).
For NISQ devices, QITE is implemented using various techniques including variational methods~\cite{Jones2019, McArdle2019, Yuan2019, Zhang2024}, probabilistic methods~\cite{Lin2021,Liu2021}, and classical determination of coefficients in the Pauli decomposition~\cite{Motta2020}. Some of these methods have been explored in several experimental studies~\cite{Motta2020, Gomes2020, Yeter2020, Sun2021, Yeter2021, Bassman2021, Yeter2022, Kamakari2022, Zhang2024}.
However, while suitable for NISQ devices, these methods face challenges in scaling up. For simulating large-sized systems, techniques designed for implementation on FTQC devices are expected to be useful. Notably, the quantum eigenvalue/singular-value transformation (QET/QSVT) technique~\cite{Gilyen2019,Martyn2021}, which incorporates qubitization~\cite{Low2019} and quantum signal processing~\cite{Low2017}, is a promising approach.
However, these techniques demand extensive quantum coherence and involve a large circuit depth, making them challenging to implement during the early stage of the FTQC era (early-FTQC era).

Here, we alternatively focus on the linear combination of unitaries (LCU) technique~\cite{Childs2012, Berry2014, Berry2015, Berry2015PRL, Low2019muliproduct}, which is notable for its independence from quantum signal processing, leading to its flexibility in quantum circuit design (see Secs.~\ref{subsec:LCU} and \ref{subsec:single-ancilla-lcu} for details).
In the LCU method, a target matrix function, such as an imaginary-time-evolution operator, is expanded in some basis where each element is implemented with a corresponding controlled-unitary gate. In the conventional implementatation, in particular, these gates are executed in succession to create a superposition of contributions of each element (see Sec.~\ref{subsec:LCU} for details).
Despite its potential, the conventional approach involves a large number of ancilla qubits and a deep quantum circuit consisting of many basic gates, making it impractical for early-FTQC devices~\cite{Campbell2022, Suzuki2022, Lin2022, Kshirsagar2022, Ding2023, Kuroiwa2023, Akahoshi2024, Akahoshi2024v2, Toshio2024}, which are limited in the number of available qubits and quantum gates. Recently, on the other hand, a single-ancilla LCU~\cite{Faehrmann2022, Chakraborty2023,Tosta2023, Wang2024, Peetz2024} has been proposed as an alternative suitable for the early-FTQC devices, involving simple quantum circuits (see Sec.~\ref{subsec:single-ancilla-lcu} for details). However, this proposal cannot be directly incorporated into calculating a finite-temperature observable since the resulting state does not correspond to the state after the QITE itself (see Sec.~\ref{sec:introduction-MCMC-SPU} for details).

In this paper, we propose a method suitable for early-FTQC devices to calculate the thermal-equilibrium expectation value of an observable at finite temperature. Our proposal, named the Markov-chain Monte Carlo with sampled pairs of unitaries (MCMC-SPU) algorithm, utilizes simple quantum circuits and overcomes the difficulties involved with the conventional LCU implementation.
Our MCMC-SPU is based on a series expansion of a QITE operator and incorporates a Markov-chain Monte-Carlo (MCMC) sampling approach similar to that used in the METTS algorithm. Initially, we perform stochastic sampling to select a pair of polynomial orders, $m$ and $n$, from the series expansion [Fig.~\ref{fig:overview-MCMCSPU}~(a)]. For each sampled order, a corresponding quantum circuit is prepared and applied to a randomly chosen input state in the computational basis [Fig.~\ref{fig:overview-MCMCSPU}~(b)]. For the output state, we measure an observable and then perform a projective measurement in the computational basis on the output state of Fig.~\ref{fig:overview-MCMCSPU}~(b), using the resulting state as the input for the next cycle [Fig.~\ref{fig:overview-MCMCSPU}~(c)]. The final step involves integrating the results of a large number of stochastic samples of pairs of orders $m$ and $n$ to calculate the thermal-equilibrium expectation value of the observable at finite temperatures [Fig.~\ref{fig:overview-MCMCSPU}~(d)]. The simplicity and small depth of each quantum circuit [Fig.~\ref{fig:overview-MCMCSPU}~(b)] make our method particularly suitable for the early-FTQC era.

The rest of this paper is organized as follows. 
Section~\ref{sec:Review} provides preliminary review of some conventional techniques for calculating the thermal-equilibrium expectation value of an observable at finite temperatures.
In Sec.~\ref{sec:introduction-MCMC-SPU}, we introduce the MCMC-SPU algorithm. 
In Sec.~\ref{sec:basic-properties}, we discuss the implementation costs associated with the MCMC-SPU algorithm, focusing on the number of qubits and quantum gates required, the probability of successful postselection, and the number of quantum circuits needed.
As an illustrative example, we present a quantitative discussion focusing on the one-dimensional transverse-field Ising model.
Section~\ref{sec:numerical-demonstration} presents a numerical demonstration of our MCMC-SPU algorithm using the same model.
Finally, we present a discussion on our proposal and conclude this paper in Sec.~\ref{sec:discussion}.

\section{Review of conventional techniques}\label{sec:Review}

In this section, we review some conventional techniques that form the basis of our method. We focus on the METTS algorithm~\cite{White2009, Stoudenmire2010}, a key approach for calculating the thermal-equilibrium expectation value of an observable for a quantum many-body system at finite temperatures.
We also examine the LCU technique~\cite{Childs2012, Berry2014, Berry2015, Berry2015PRL, Low2019muliproduct} as a specific method for implementing imaginary-time evolution, which is vital to the METTS algorithm.
Furthermore, we discuss the technique of the single-ancilla LCU~\cite{Faehrmann2022, Chakraborty2023, Tosta2023, Wang2024, Peetz2024}, a recent adaptation of the LCU algorithm that is well-suited for early-FTQC devices with a limited number of qubits and quantum gates.

\subsection{METTS algorithm}\label{subsec:METTS}

A prominent approach for calculating the thermal-equilibrium expectation value of an observable for a quantum many-body system at finite temperatures is the METTS algorithm~\cite{White2009, Stoudenmire2010}.
In the standard framework of statistical mechanics, the thermal-equilibrium state of a system coupled to a thermal bath is described by the canonical statistical ensemble, or the Gibbs state. For this state, the canonical ensemble average of an observable $O$ is given by
\begin{align} \label{eq:canonical_average}
\ev{O}_\beta
= \frac{1}{Z} \Tr\qty[O e^{-\beta H}],
\end{align}
where $\beta$ denotes the inverse temperature of the thermal bath that is coupled to the target system,
and $Z$ denotes the partition function, as follows:
\begin{align} \label{eq:partition-function}
Z = \Tr\qty[ e^{-\beta H}].
\end{align}
In the computational basis $\{\ket{i}\}$, which consists of the classical product states (CPSs) $\ket{i}$'s, the canonical ensemble average in Eq.~\eqref{eq:canonical_average} for an $N$-qubit system can be rewritten as follows:
\begin{align}
\ev{O}_\beta
&= \frac{1}{Z} \sum_{i=0}^{2^N -1} \bra{i} e^{-(\beta/2) H} O \, e^{-(\beta/2) H} \ket{i} \nonumber\\
&=  \sum_{i=0}^{2^N -1} \frac{P_i}{Z} \ev{O}{\phi_i}, \label{eq:canonical-expectation-METTS}
\end{align}
where the normalized quantum state $\ket{\phi_i}$ corresponding to the CPS $\ket{i}$ is called the METTS~\cite{White2009, Stoudenmire2010} with definition
\begin{align}
\ket{\phi_i}
:= \frac{1}{P_i^{1/2}} e^{-(\beta/2) H} \ket{i} \label{eq:METTS-def}
\end{align}
and the normalization factor
\begin{align}
P_i := \ev{e^{-\beta H}}{i}.
\end{align}
The set of all the normalization factors $P_i$'s satisfy the following relationship [see Eq.~\eqref{eq:partition-function}]:
\begin{align}
Z = \sum_{i=0}^{2^N -1} P_i.
\end{align}
To calculate the canonical ensemble average of an observable $O$ as seen in Eq.~\eqref{eq:canonical-expectation-METTS}, we use the ensemble average across a set of the METTS $\ket{\phi_{i}}$'s, with a probability distribution $\{P_{i}/Z \}$.
The METTS algorithm~\cite{White2009, Stoudenmire2010} generates this statistical ensemble as a stationary distribution through MCMC sampling~\cite{Geyer1992, Gilks1995, Cowles1996, Brooks1998}.
In the following, we present the concrete procedure of the METTS algorithm:
\begin{enumerate}
	\item Randomly selecting a CPS $\ket{i}$ from the computational basis.

    \item Performing an imaginary-time evolution to the CPS $\ket{i}$ to obtain the corresponding METTS $\ket{\phi_{i}}$ as the resulting normalized state:
        \begin{align} \label{eq:second-step-METTS}
            \ket{i} \to \ket{\phi_i} = \frac{1}{P_i^{1/2}} e^{-(\beta/2) H} \ket{i}.
        \end{align}

    \item Measuring an observable $O$ with respect to the METTS $\ket{\phi_i}$ to obtain its quantum mechanical expectation value
        \begin{align} \label{eq:third-step-METTS}
            \ev{O}_i = \ev{O}{\phi_i}.
        \end{align}

    \item Collapsing the METTS $\ket{\phi_i}$ onto some CPS $\ket{j}$ in the computational basis, i.e., 
        \begin{align}
            \ket{\phi_i} \to \ket{j}
        \end{align}
        according to the probability $p(i\to j)$ given by the squared overlap between these states:
        \begin{align} \label{eq:transision-probability-METTS}
            \hspace{7mm} p(i\to j)=\qty|\braket{j}{\phi_i}|^2 = \frac{1}{P_i} \qty|\mel{j}{e^{-(\beta/2)H}}{i}|^2.
        \end{align}
        	The resulting state $\ket{j}$ serves as the input state for Step~2 [see Eq.~\eqref{eq:second-step-METTS}] in the next cycle, which is followed by Steps~3 and 4.
        	
	\item Calculating the statistical mean of the quantum mechanical expectation values [see Eq.~\eqref{eq:third-step-METTS}] after many cycle repetitions consisting of Steps~2--4.
\end{enumerate}
In particular, the quantum mechanical expectation value~\eqref{eq:third-step-METTS} in Step~3 is obtained using an ensemble consisting of many copies of $\ket{\phi_i}$, and Step~4 is executed on an additional copy of that state.
Originally, the METTS algorithm was developed for classical computers, but a similar approach can be adopted for quantum computers, known as the QMETTS algorithm, as mentioned in Refs.~\cite{Motta2020, Sun2021, Bassman2021, Getelina2023, Pollock2023, Gacon2024, Shen2024, Qian2024, Summer2024}.
In the QMETTS algorithm, Step~2 in the procedure is performed using QITE on a quantum computer, while Step~4 is achieved through projective measurement in the computational basis.

The target distribution $\{ P_i / Z \}$ satisfies the following relationship:
\begin{align}
\sum_{i=0}^{2^N -1} \frac{P_i}{Z} \, p(i \to j)
&= \sum_{i=0}^{2^N -1} \frac{P_i}{Z} \frac{1}{P_i} \qty|\mel{j}{e^{-(\beta/2)H}}{i}|^2 \nonumber \\
&= \frac{P_j}{Z}.
\end{align}
Therefore, if the process is irreducible, aperiodic, and positive, the target distribution is the stationary distribution of the Markov process $i \to j$ defined in Step~4. 
Once the Markov process has reached convergence, calculating the statistical average of the expectation values~\eqref{eq:third-step-METTS} will yield the desired finite-temperature expectation value.


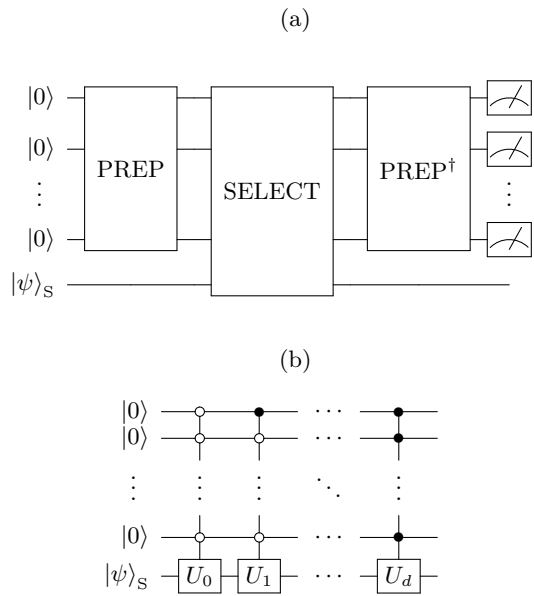
\begin{figure}[htbp]
    \centerline{{(a)} }
    \[
\Qcircuit @C=0.7em @R=0.7em {
\lstick{\ket{0}} & \multigate{3}{\rm PREP} & \qw  &\multigate{4}{\rm SELECT} & \qw& \multigate{3}{\rm PREP^{\dag}} & \meter \\
\lstick{\ket{0}} & \ghost{\rm PREP} &\qw &\ghost{\rm SELECT} &\qw & \ghost{\rm PREP^{\dag}} & \meter \\
\lstick{\raisebox{6pt}{\vdots}~\,} & \pureghost{\rm PREP} & & \pureghost{\rm SELECT} &&\pureghost{\rm PREP^\dag} & \raisebox{6pt}{\vdots} \\
\lstick{\ket{0}} & \ghost{\rm PREP} &\qw &\ghost{\rm SELECT} &\qw &\ghost{\rm PREP^{\dag}} & \meter \\
\lstick{\ket{\psi}_{\rm S}} & \qw & \qw& \ghost{\rm SELECT} & \qw &\qw &\qw
}
\]
    \vspace{0.25cm}
    \centerline{{(b)} }
    \[
\Qcircuit @C=0.7em @R=0.7em {
\lstick{\ket{0}} &  \ctrlo{1} & \ctrl{1} & \qw & \cdots & & \ctrl{1}&\qw \\
\lstick{\ket{0}} & \ctrlo{1} & \ctrlo{1} & \qw & \cdots & & \ctrl{1} &\qw \\
\lstick{\raisebox{6pt}{\ensuremath{\vdots}}~\,} & \push{\raisebox{6pt}{\ensuremath{\vdots}}} & \push{\raisebox{6pt}{\ensuremath{\vdots}}} & & \push{\raisebox{6pt}{\ensuremath{\ddots}}} && \push{\raisebox{6pt}{\ensuremath{\vdots}}} \\
\lstick{\ket{0}} &  \ctrlo{-1} \qwx[1] & \ctrlo{-1} \qwx[1] & \qw & \cdots & & \ctrl{-1} \qwx[1] & \qw \\
\lstick{\ket{\psi}_{\rm S}} &  \gate{U_{0}} & \gate{U_{1} } & \qw & \cdots & & \gate{U_{d} } & \qw 
}
\]
    \caption{
    Quantum circuit for the conventional form of the LCU technique.
    (a) A set of coefficients $\{c_k\}$ appearing in  Eq.~\eqref{eq:LCU-general} is encoded in the ancilla qubits at the PREPARE circuit [see Eq.~\eqref{eq:PREPARE}], labeled ``{\rm PREP}" in the figure.
    A series of unitary gates $\{U_{k} \}$ appearing in Eq.~\eqref{eq:LCU-general} is applied to the target state $\ket{\psi}_{\rm S}$ successively at the SELECT circuit [see Eq.~\eqref{eq:LCU-SELECT}], which is labeled ``{\rm SELECT}" in the figure.
    (b) Prototypical example of the implementation of the SELECT circuit, which takes the form of a series of multicontrolled unitary gates. The implementation can potentially be simplified depending on the specific form of $\{U_k\}$.
    }
    \label{fig:coherentLCU}
\end{figure}

\subsection{Linear combination of unitaries}\label{subsec:LCU}

As discussed in the previous subsection, implementing an imaginary-time evolution on a quantum device plays a crucial role in the QMETTS algorithm [see Eq.~\eqref{eq:second-step-METTS}].
Among the various methods of implementation proposed so far, this paper emphasizes the LCU technique~\cite{Childs2012, Berry2014, Berry2015, Berry2015PRL, Low2019muliproduct}, chosen for its independence from quantum signal processing~\cite{Low2017} and its flexibility in quantum-circuit design.
In the following, we provide a brief overview of this technique.
The LCU method is suitable for implementation on FTQC devices and is useful for implementing a variety of matrix functions $f(M)$, where the argument is some matrix $M$. To implement such a matrix function on a quantum circuit, as shown in Fig.~\ref{fig:coherentLCU},
we approximate the target function $f(M)$ using a linear combination of $(d+1)$ terms of simple unitary matrices $U_k (M)$ as follows:
\begin{align} \label{eq:LCU-general}
f(M) \simeq \tilde{f}(M) = \sum_{k=0}^d c_k \, U_k (M),
\end{align}
where each unitary matrix $U_k (M)$ is a function of the matrix $M$, and
$c_k$ denotes the corresponding real positive coefficient.
A set of coefficients $\{c_k\}$ is encoded in ancilla qubits, initialized as a product of $\ket{0}$ states, as shown in Fig.~\ref{fig:coherentLCU}. Each unitary matrix $U_k (M)$ is successively applied to the target system, represented as $\ket{\psi}_{\rm S}$ in Fig.~\ref{fig:coherentLCU}.

The encoding of the set of coefficients $\{c_k\}$ appearing in Eq.~\eqref{eq:LCU-general} into the ancilla qubits is implemented at the PREPARE circuit~\cite{Childs2012}, represented as the ``PREP" box in Fig.~\ref{fig:coherentLCU}(a). The action of the PREPARE circuit is given by
\begin{align} \label{eq:PREPARE}
{\rm PREP} \ket{0}_{\! \rm A}
= \sum_{k=0}^d \sqrt{\frac{c_k}{\sum_{k'=0}^d \, c_{k'}}} \ket{k}_{\! \rm A},
\end{align}
where $\ket{0}_{\! \rm A}:= \ket{0}^{\otimes \lceil \log_2 (d+1) \rceil}$ and $\ket{k}_{\! \rm A}$'s are multi-qubit ancilla states in the computational basis.
Subsequently, a successive application of a series of unitary matrices $\{U_k (M) \}$ to the target state $\ket{\psi}_{\rm S}$ is implemented at the SELECT circuit~\cite{Childs2012}. Specifically, a unitary gate $U_k (M)$ is applied to the target state $\ket{\psi}_{\rm S}$ under the condition that the ancilla qubits are in the $\ket{k}_{\! \rm A}$ state (see Fig.~\ref{fig:coherentLCU}), i.e.,
\begin{align} \label{eq:LCU-SELECT}
{\rm SELECT} = \sum_{k=0}^d \ket{k}_{\!\rm A \, A}\!\!\bra{k}\otimes U_k (M).
\end{align}
A prototypical example of implementing the SELECT circuit is shown in Fig.~\ref{fig:coherentLCU}(b), which takes the form of a series of multicontrolled unitary gates. The implementation can potentially be simplified depending on the specific form of $\{U_k\}$.
Combining the SELECT circuit with the PREPARE circuit and its Hermitian conjugate, we approximately realize the desired matrix function $f(M) \, [\simeq \tilde{f}(M)]$ up to an overall constant factor as follows [see Fig.~\ref{fig:coherentLCU}(a)]:
\begin{align}
&\qty(\,_{\rm A \!}\!\bra{0} {\rm PREP}^{\dag} ) \cdot {\rm SELECT} \cdot ({\rm PREP} \ket{0}_{\! \rm A}\otimes \ket{\psi}_{\rm S} ) \nonumber
 \\
&= \frac{\tilde{f}(M)}{\sum_{k'=0}^d \, c_{k'}} \ket{\psi}_{\rm S}, \label{eq:PREP-SELSECT-PREP}
\end{align}
where the ancilla qubits are initialized as the $\ket{0}_{\! \rm A}$ state.
In Eq.~\eqref{eq:PREP-SELSECT-PREP}, after applying the $\rm PREP^\dag$ operation, we perform a measurement on the ancilla qubits and make a postselection of the events where the measurement outcome is $\ket{0}_{\! \rm A}$.
This postselection effectively restricts the quantum circuit's action to a specific block element of the representation matrix, which is a method known as block encoding~\cite{Childs2017, Loke2017, Gilyen2019, Camps2022, Camps2023}.

In the rest of Sec.~\ref{subsec:LCU}, we discuss an LCU implementation of an imaginary-time evolution, which plays a crucial role in the QMETTS algorithm (see Sec.~\ref{subsec:METTS}).

\subsubsection{LCU implementation of an imaginary-time evolution}\label{subsubsec:LCU-QITE}

We first perform a decomposition of the imaginary-time-evolution operator. We employ an expansion based on a series of the Chebyshev polynomials $\{T_k (\cdot)\}$ as follows~\cite{Abramowitz1964, Tosta2023, Marteau2023}:
\begin{align}
e^{-(\beta/2) H}
\simeq I_0 ( \beta/2) \, I 
+ 2 \sum_{k=1}^d I_k (\beta/2) \, T_k (-H),
\label{eq:expansion-QITE}
\end{align} 
where $I_k (x)$ denotes the $k$-th order modified Bessel function of the first kind, which satisfies $I_k(x)>0$ for $x>0$, and the expansion is truncated up to the order $d$.
In Eq.~\eqref{eq:expansion-QITE}, the Hamiltonian $H$ is normalized so that its eigenvalues do not exceed one in absolute value.
In the LCU implementation, 
each matrix function $T_k (-H)$ corresponds to $U_k (M)$ in the general form in Eq.~\eqref{eq:LCU-general}. The specific implementation features block encoding of the $k$-times iteration $\qty(W_H)^k$ of the walk operator $W_H$~\cite{Low2019,Marteau2023} combined with conditioning on the state of the ancilla qubits, i.e., [see Eq.~\eqref{eq:PREP-SELSECT-PREP}]
\begin{align}
&\qty(\,_{\rm A \!}\!\bra{0} {\rm PREP}^{\dag} ) \, \qty(W_H)^k \, ({\rm PREP} \ket{0}_{\! \rm A}\otimes \ket{\psi}_{\rm S} ) \nonumber
 \\
&= T_k (-H) \ket{\psi}_{\rm S}.
\label{eq:modified_Bessel_block_encoding}
\end{align}
The walk operator $W_H$ itself can also be constructed using the LCU structure for the block encoding of the Hamiltonian $H$. 
Specifically, when the SELECT circuit ${\rm SELECT}_H$ used for the block encoding of the Hamiltonian satisfies $\qty({\rm SELECT}_H)^2 = I$, the walk operator can be obtained as follows~\cite{Low2019,Marteau2023}:
\begin{align} \label{eq:qubitization-walk-operator}
W_H = S_H \cdot {\rm SELECT}_H
\end{align}
using the reflection $S_H$ with respect to the ancilla state ${\rm PREP}_H \ket{0}_{{\rm A}, H}$, i.e., 
\begin{align}\label{eq:reflection}
S_H = {\rm PREP}_H\qty(2 \ket{0}_{{\rm A}, H} \,_{{\rm A}, H}\bra{0}  - I_{\rm A} ) {{\rm PREP}}^\dag_H \otimes I_{\rm S}. 
\end{align}
Here, ${\rm PREP}_H$ denotes the PREPARE circuit for the block encoding of the Hamiltonian, and $\ket{0}_{{\rm A}, H}$ denotes the product of $\ket{0}$ states of the corresponding ancilla qubits.

\subsubsection{Successful postselection probability for quantum imaginary-time evolution}\label{subsubsec:PSprob-QITE}

As discussed in Sec.~\ref{subsec:LCU}, using the LCU technique for QITE involves measuring the ancilla qubits and postselection of specific outcomes (see Eq.~\eqref{eq:PREP-SELSECT-PREP} and the following argument).
Generally, the probability of successful postselection decays exponentially with increasing imaginary time or inverse temperature $\beta$~\cite{Poulin2009, Chowdhury2017, vanApeldoorn2020, Silva2023}.

Specifically, for an initial state $\ket{\psi} = \sum_j \tilde{c}_j \ket{E_j}$, where $\ket{E_j}$ denotes the energy eigenstate with eigenenergy $E_j$,
the unnormalized vector obtained as a result of a successful postselection for the LCU implementation of an imaginary-time evolution is determined as follows [see Eqs.~\eqref{eq:PREP-SELSECT-PREP} and \eqref{eq:expansion-QITE}]:
 \begin{align} \label{eq:QITE-unnormalized}
 \ket{\psi(\beta/2)}
 &\simeq \frac{e^{-(\beta/2)H}}{e^{\beta/2}} \sum_j \tilde{c}_j \ket{E_j} \nonumber\\
 &= \sum_j \tilde{c}_j e^{-(\beta/2)(E_j - E_{\rm GS})} \ket{E_j}.
 \end{align}
Here, we assume that the eigenenergies are normalized such that $-1 = E_{\rm GS} < E_1 < \cdots < E_j <\cdots\le 1$, which aligns with the expansion using the Chebyshev polynomials as stated in Eq.~\eqref{eq:expansion-QITE}.
In Eq.~\eqref{eq:QITE-unnormalized}, we have also used the approximation $I_0 (\beta/2) + 2 \sum_{k=1}^d I_k (\beta/2) \simeq e^{\beta/2}$ for a sufficiently large value of $d$~\cite{Tosta2023} (see Eqs.~\eqref{eq:error-fA} and \eqref{eq:bound_c_norm} for details).
The probability of a successful postselection of the measurement outcome on the ancilla qubits is proportional to the squared norm of the unnormalized vector in Eq.~\eqref{eq:QITE-unnormalized}, which satisfies the following condition:
\begin{align} \label{eq:PSprob_generalLCU}
p_{\rm PS} \propto  \braket{\psi(\beta/2)}
\ge e^{-2\beta}.
\end{align}
In this evaluation, we have used the relationship $ \braket{\psi(\beta/2)} \simeq \sum_j |\tilde{c}_j|^2 e^{-\beta(E_j +1)} $ together with the normalization condition $\sum_j |\tilde{c}_j|^2 =1$.
From Eq.~\eqref{eq:PSprob_generalLCU}, the number of trials $N_{\rm PS}$ required to ensure that we achieve a successful measurement outcome for general models scales as $N_{\rm PS} = \mathcal{O}\qty(e^{2\beta})$ in the worst-case scenario (see also Ref.~\cite{Chowdhury2021}).

\subsection{Single-ancilla LCU}\label{subsec:single-ancilla-lcu}

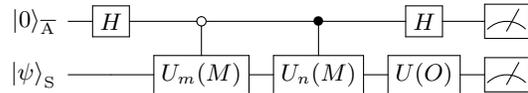
\begin{figure}[tbp]
    \[
    	\Qcircuit @C=1em @R=.7em {
  	\lstick{\ket{0}_{ \overline{\rm A}}} & \gate{H} & \ctrlo{1} & \ctrl{1}    & \gate{H}      & \meter \\
  	\lstick{\ket{\psi}_{\rm S}} & \qw      & \gate{ U_{m} (M)}    & \gate{ U_{n} (M)}       & \gate{U(O)}    & \meter 
	}
    \]
    \caption{
    Quantum circuit for the single-ancilla LCU.
    The circuit consists of a pair of unitary gates $U_m (M)$ and $U_n (M)$, which are stochastically sampled from the set of the unitary gates $\{U_0 (M), U_1 (M), \cdots, U_d (M)\}$ appearing in the conventional LCU method (see Fig.~\ref{fig:coherentLCU}).
    Unlike the conventional approach, this implementation uses only a single ancilla qubit, which serves as the control degree of freedom for the controlled unitary gates.
    After applying the circuit to the initial state $\ket{0}_{ \overline{\rm A}}\otimes \ket{\psi}_{\rm S}$, 
    we simultaneously measure the Pauli $Z$ operator of the ancilla qubit and a general observable $O$ for the target system, the latter of which is implemented using a unitary gate $U(O)$.
    By calculating the statistical mean of the results for various pairs of unitary gates that are stochastically sampled, we can reproduce the result of the original LCU method.
    }
    \label{fig:singleLCU}
\end{figure}

The conventional LCU algorithm discussed in Sec.~\ref{subsec:LCU} involves many ancilla qubits and a deep quantum circuit consisting of a large number of basic gates, making it unsuitable for the early stage of the FTQC regime (see Figs.~\ref{fig:resource_implementation}(a)--(c) for an illustrative example). Recently, a variant of the LCU method has been proposed~\cite{Faehrmann2022, Chakraborty2023, Tosta2023, Wang2024, Peetz2024}, featuring simpler quantum circuits, as shown in Fig.~\ref{fig:singleLCU}, though it involves a statistical sampling overhead (see also Sec.~\ref{subsec:number-circuits}).
This variant, referred to as ``single-ancilla LCU" in Ref.~\cite{Chakraborty2023}, is named after its quantum-circuit structure. In the following, we present a brief review of this method.

We focus on a certain form~\cite{Faehrmann2022, Chakraborty2023,Tosta2023, Wang2024, Peetz2024} of the single-ancilla LCU, which can be used to estimate the expectation value $\ev{O}_{f(M)}$ of an observable $O$ with respect to the resulting unnormalized state $f(M) \ket{\psi}_{\rm S}$ after applying a matrix function $f(M)$ to the initial state $\ket{\psi}_{\rm S}$, i.e.,
\begin{align}
\ev{O}_{f(M)} = \ev{f(M) \, O \, f(M)}{\psi}.
\end{align}
Applying the LCU decomposition in Eq.~\eqref{eq:LCU-general} to this equation, $\ev{O}_{f(M)}$ is approximated as follows:
\begin{align}
&\ev{O}_{\tilde{f}(M)}
= \sum_{m=0}^d \sum_{n=0}^d
 c_m c_n \,_{\rm S}\!\bra{\psi} U_m^\dag (M) \, O \, U_n (M) \ket{\psi}_{\rm S} \nonumber\\
&=  \sum_{m=0}^d \sum_{n=0}^d p_m p_n ||c||_1^2 \,_{\rm S}\!\bra{\psi} U_m^\dag (M) \, O \, U_n (M) \ket{\psi}_{\rm S}, \label{eq:single_ancilla_LCU}
\end{align}
where we have introduced the following notations: $||c||_1 = \sum_{n=0}^d c_n$ and $p_n = c_n / ||c||_1$.
Here, for the $d$-th-order approximation $\tilde{f}(\cdot)$ of $f(\cdot)$ satisfying \begin{align} \label{eq:error-fA}
\max_{x \in {\rm Dom}(f)} | f(x) - \tilde{f}(x) | \le \nu
\end{align}
with the domain ${\rm Dom}(f)$ of $f(\cdot)$,
the approximation error in the observable is upper bounded
as follows~\cite{Tosta2023}
\begin{align} \label{eq:error-observable}
| \ev{O}_{\tilde{f}(M)} - \ev{O}_{f(M)} | \le \nu ||O||_{\rm op} \qty(2||f(M) ||_{\rm op} + \nu)
\end{align}
with operator norm $|| \cdot ||_{\rm op}$.
The quantity on the right-hand side of Eq.~\eqref{eq:single_ancilla_LCU} can be obtained by stochastically sampling a pair of unitary gates $U_m$ and $U_n$ according to the probability distribution $\{p_n\}$ and by calculating the value of a stochastic variable $||c||_1^2 \,_{\rm S}\!\bra{\psi} U_m^\dag (M) \, O \, U_n (M) \ket{\psi}_{\rm S}$. The term $ \,_{\rm S}\!\bra{\psi} U_m^\dag (M) \, O \, U_n (M) \ket{\psi}_{\rm S}$ is obtained using the quantum circuit shown in Fig.~\ref{fig:singleLCU}, which consists of the stochastically sampled pair of unitary gates $U_m (M)$ and $U_n (M)$.
Unlike the conventional implementation (see Fig.~\ref{fig:coherentLCU}), this approach uses only a single ancilla qubit.
The state after applying the quantum circuit to the initial state $\ket{0}_{\overline{\rm A}}\otimes\ket{\psi}_{\rm S}$ is determined as follows:
\begin{align} \label{eq:result-singleLCU}
\ket{0}_{\overline{\rm A}} \otimes \frac{1}{2}(U_m + U_n)\ket{\psi}_{\rm S}
+ \ket{1}_{\overline{\rm A}} \otimes \frac{1}{2}(U_m - U_n)\ket{\psi}_{\rm S}.
\end{align}
For this state, we evaluate the expectation value $\ev{O}_{m n}^0$ ($\ev{O}_{m n}^1$) of an observable $O$ for the target system under the condition that the outcome of a simultaneous measurement on the ancilla qubit is $0$ ($1$), as follows:
\begin{align}
\ev{O}_{m n}^0
 &= \frac{1}{4} \,_{\rm S}\!\bra{\psi} (U_m^\dag + U_n^\dag)\, O \, (U_m + U_n) \ket{\psi}_{\rm S},\\
 \ev{O}_{m n}^1
 &= \frac{1}{4} \,_{\rm S}\!\bra{\psi} (U_m^\dag - U_n^\dag) \, O \, (U_m - U_n) \ket{\psi}_{\rm S}.
\end{align}
From these equations, we obtain the desired quantity as follows:
\begin{align}
\ev{O}_{m n}^0 - \ev{O}_{m n}^1
&= \frac{1}{2} \,_{\rm S}\!\bra{\psi} \qty( U_m^\dag O U_n + U_n^\dag O U_m ) \ket{\psi}_{\rm S} \nonumber\\
&= \Re\qty[\,_{\rm S}\!\bra{\psi}  U_m^\dag O U_n \ket{\psi}_{\rm S}],
\end{align}
which is sufficient to evaluate the value on the right-hand side of Eq.~\eqref{eq:single_ancilla_LCU}, since a stochastic sampling of a pair $(m, n)$ and that of $(n, m)$ are equivalent; thus, only the real part of each term contributes to the summation.

\section{Markov-chain Monte Carlo with sampled pairs of unitaries}\label{sec:introduction-MCMC-SPU}

As discussed in the previous section, the single-ancilla LCU is suitable for the early-FTQC era, particularly for implementing a QITE (see Sec.~\ref{subsec:single-ancilla-lcu}), which plays a crucial role in simulating finite-temperature quantum many-body systems (see Sec.~\ref{subsec:METTS}).
However, the state obtained after applying the quantum circuit in the single-ancilla LCU [see Eq.~\eqref{eq:result-singleLCU}] does not match the METTS itself described in Eq.~\eqref{eq:METTS-def}, indicating that QITE in the QMETTS algorithm cannot be replaced with the single-ancilla LCU in a straightforward manner.

In this section, we propose the MCMC-SPU algorithm (see Fig.~\ref{fig:overview-MCMCSPU}) designed for the early-FTQC era. This algorithm calculates the thermal-equilibrium expectation value of an observable at finite temperatures. Our proposal utilizes a type of MCMC sampling to generate a statistical ensemble corresponding to each quantum circuit (see Fig.~\ref{fig:circuit-MCMC-SPU}) used in the single-ancilla-LCU implementation of QITE.

\subsection{Procedure of the MCMC-SPU algorithm}\label{subsec:MCMC-SPU-procedure}

In the following, we present a detailed procedure for our MCMC-SPU algorithm, which is built upon QITE [see Eq.~\eqref{eq:expansion-QITE}] implemented with the single-ancilla-LCU method (see Sec.~\ref{subsec:single-ancilla-lcu}).
\begin{figure}[tbp]
    \centerline{{(a)} }
   \[
    	\Qcircuit @C=1em @R=.7em {
  	\lstick{\ket{0}_{ \overline{\rm A}}} & \gate{H} & \multigate{2}{U_{m n}}    & \gate{H}      & \meter \\
  	\lstick{\ket{0}^{\otimes \lceil \log_2 d' \rceil}_{\! \rm A}} & \gate{{\rm PREP}_H}      & \ghost{ U_{m n}}    &\gate{{\rm PREP}_H^\dag}&\meter \\
  	\lstick{\ket{i}_{\rm S}} & \qw      & \ghost{U_{m n}} & \qw  &\qw
	}
    \]
    \vspace{0.25cm}
    \centerline{{(b)} }
    \[
    \Qcircuit @C=1em @R=.7em {
  	&\qw & \ctrlo{1} & \ctrl{1} &\qw    \\
  	& \qw      & \multigate{1}{ \qty(W_H)^{m}}    & \multigate{1}{ \qty(W_H)^{n}} &\qw \\
  	& \qw      & \ghost{\qty(W_H)^{m}} & \ghost{\qty(W_H)^{n}} & \qw  
}
    \]

    \caption{
    (a) Quantum circuit for the proposed MCMC-SPU. The specific form of the gate $U_{m n}$ is provided in (b), which corresponds to a sampled pair of orders $m$ and $n$.
    In Fig.~(b), each box represents the $m$($n$)-times iteration $\qty(W_H)^{m (n)}$ of the walk operator $W_H$ [see Eq.~\eqref{eq:qubitization-walk-operator}] forming the block encoding of the $m$($n$)-th-order Chebyshev polynomial.
    The ancilla state $\ket{G}_{\! \rm A} = {\rm PREP}_H \ket{0}_{\! \rm A}^{\otimes \lceil \log_2 d' \rceil}$
    is prepared using the PREPARE circuit ${\rm PREP}_H$ for the block encoding of the Hamiltonian $H$ consisting of $d'$ terms, while $\ket{i}_{\rm S}$ denotes the target system in the computational basis.
    }
    \label{fig:circuit-MCMC-SPU}
\end{figure}
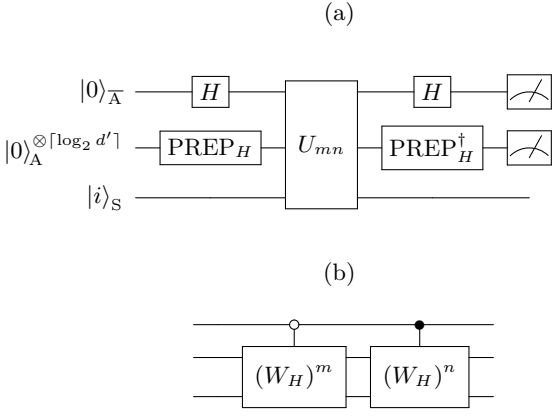
%
\begin{enumerate}

    \item Sampling a pair of orders $m$ and $n$ of the Chebychev polynomials in Eq.~\eqref{eq:expansion-QITE} in a stochastic manner according to the probability distribution $\{p_n = c_n / (\sum_{n'=0}^d c_{n'}) \}$, where the coefficient $c_n$ is given by
    \begin{align}
    c_n = 
    \begin{cases}
        I_0 (\beta/2) & (n=0);\\
        2 I_n (\beta/2) & (n \ne 0).
    \end{cases} \label{eq:MCMC-SPU-coefficient}
    \end{align}

    \item Constructing a quantum circuit $U_{m n}$ as shown in Fig.~\ref{fig:circuit-MCMC-SPU}(b), which corresponds to the sampled pair of orders $m$ and $n$.
    In Fig.~\ref{fig:circuit-MCMC-SPU}(b), $\qty(W_H)^{m (n)}$ denotes the $m$($n$)-times iteration of the walk operator $W_H$ [see Eq.~\eqref{eq:qubitization-walk-operator}] forming the block encoding of the $m$($n$)-th order Chebyshev polynomial.

\item Preparing an initial state
 \begin{align} \label{eq:MCMC-SPU-initial-state}
    \ket{i}_{\rm tot} :=
     \ket{0}_{\overline{\rm A}} \otimes \ket{G}_{\! \rm A}\otimes \ket{i}_{\rm S},
 \end{align}
 where $\ket{i}_{\rm S}$ denotes a CPS randomly chosen from the computational basis for the target system.
 Here, $\ket{0}_{\overline{\rm A}}$ denotes an ancilla qubit that acts as the control qubit for $\qty(W_H)^{m}$ and $\qty(W_H)^{n}$, while the ancilla state $\ket{G}_{\! \rm A} = {\rm PREP}_H \ket{0}_{\! \rm A}^{\otimes \lceil \log_2 d' \rceil}$
    is prepared with ${\rm PREP}_H$ [see Fig.~\ref{fig:circuit-MCMC-SPU}(a)], which is a specific form of the PREPARE circuit (see Eq.~\eqref{eq:PREPARE} for the general form) for the block encoding of the Hamiltonian $H$.
    
    \item Applying the quantum circuit $U_{m n}$ shown in Fig.~\ref{fig:circuit-MCMC-SPU}(b) to the initial state in Eq.~\eqref{eq:MCMC-SPU-initial-state}
    and then applying ${\rm PREP}_H^\dag$ to the ancilla qubits. 
    
    \item Performing measurement on the ancilla qubits in the computational basis and making postselection of measurement outcome of the ancilla qubits with the result $\ket{k}_{\overline{\rm A}} \otimes \ket{0}_{\! \rm A}^{\otimes \lceil \log_2 d' \rceil}$ for $k \in \{0, 1\}$, which results in projection onto a normalized state $\ket{\phi_{m n}^{i k}}$ (see Eq.~\eqref{eq:generalizedMETTS} in Sec.~\ref{subsec:MCMC-SPU-theory} for details).
        
    \item Measuring an observable $O$ with respect to the resulting state $\ket{\phi_{m n}^{i k}}$ to obtain its quantum mechanical expectation value as follows:
        \begin{align} \label{eq:ev-MCMC-SPU}
            \ev{O}_{m n}^{i k} = \ev{O}{\phi_{m n}^{i k}}.
        \end{align}
    Evaluating this expectation value is executed with an ensemble consisting of many copies of $\ket{\phi_{m n}^{i k}}$.
    
    \item Collapsing an additional copy of the state $\ket{\phi_{m n}^{i k}}$ onto some CPS $\ket{j}_{\rm S}$ in the computational basis for the target system by performing projective measurement on the target system. 
    A transition $\ket{\phi_{m n}^{i k}} \to \ket{j}_{\rm S}$ occurs with a probability $\qty|_{\rm {S}}\!\braket{j}{\phi_{m n}^{i k}}|^2$ (see Eq.~\eqref{eq:MCMC-SPU_transition} in Sec.~\ref{subsec:MCMC-SPU-theory} for details).
    The resulting state $\ket{j}_{\rm S}$ serves as the input state for Step~3 [see Eq.~\eqref{eq:MCMC-SPU-initial-state}] in the next cycle, which is followed by Steps~4--6.
        	
    \item Calculating the statistical mean $\ev{O}_{m n}^k$ (see Eq.~\eqref{eq:MCMC-SPU-each-ensemble-average} for the definition) of the quantum mechanical expectation values in Eq.~\eqref{eq:ev-MCMC-SPU} after a large number of cycle repetitions consisting of Steps~3--7.

    \item Repeating Steps~1--8 many times with resampling a pair of orders $m$ and $n$ for each cycle, and then calculate the statistical ensemble average of $\{\ev{O}_{m n}^k\}$ with an appropriately defined weight (see Eqs.~\eqref{eq:MCMC-SPU-total-result}--\eqref{eq:MCMC-SPU-each-ensemble-average} in the following).
    
\end{enumerate}

\subsection{Validity of the MCMC-SPU algorithm}\label{subsec:MCMC-SPU-theory}

We now present the proof of the validity of the MCMC-SPU algorithm.
In Step~5 of Sec.~\ref{subsec:MCMC-SPU-procedure}, the resulting unnormalized vector after measuring the ancilla qubits is provided as follows (see Fig.~\ref{fig:circuit-MCMC-SPU}):
\begin{align}
&\qty(\bra{k}_{\rm \overline{A}} \otimes \bra{G}_{\rm A}) U_{m n} \ket{i}_{\rm tot} \nonumber \\
= &\frac{1}{\sqrt{2}} \qty[T_m (-H) + (-1)^k \, T_n (-H)] \ket{i}_{\rm S},\label{eq:single_ancilla_Chebyshev}
\end{align} 
from which we obtain the corresponding normalized state vector
\begin{align}
\ket{\phi_{m n}^{i k}}
:= \frac{1}{\sqrt{W_{m n}^{i k}}}
\qty(\bra{k}_{\rm \overline{A}} \otimes \bra{G}_{\rm A}) U_{m n} \ket{i}_{\rm tot} \label{eq:generalizedMETTS}
\end{align}
with a probability of successful postselection, which is determined as follows:
\begin{align}
\label{eq:MCMC-SPU-PSprob}
W_{m n}^{i k}
:= \,_{\rm tot}\!\bra{i}U^\dag_{m n}\qty(\ketbra{k}_{\rm \overline{A}} \otimes \ketbra{G}_{\rm A}\otimes I_{\rm S}) U_{m n} \ket{i}_{\rm tot}.
\end{align}
The value of $W_{m n}^{i k}$ is real and nonnegative since it represents the expectation value of a projection operator.
In particular, it follows from Eq.~\eqref{eq:single_ancilla_Chebyshev} that
\begin{align} \label{eq:W_no_contribution}
\qty(\bra{1}_{\rm \overline{A}} \otimes \bra{G}_{\rm A}) U_{m m} \ket{i}_{\rm tot} = 0.
\end{align}
This implies that for the pair of orders $(m, n)$ with $m=n$ and $k=1$ has no contribution to the finite-temperature observable. As a result, we skip Steps 2--8 for these index combinations.
Based on the relationship in Eq.~\eqref{eq:single_ancilla_Chebyshev}, the thermal expectation value of an observable $O$ can be expressed in terms of its quantum mechanical expectation value for the normalized states $\ket{\phi_{m n}^{i k}}$'s as follows:
\begin{align}
&\ev{O}_{\beta}
\simeq \frac{ \sum_{m, n=0}^d 
\sum_{i} c_m c_n  \,_{\rm S \!\!}\ev{T_m (-H) \, O \, T_n (-H)}{i}_{\rm S}
}{
\sum_{m',n'=0}^d \sum_{i'} c_{m'} c_{n'}  \,_{\rm S \!\!}\ev{T_{m'} (-H) \, T_{n'} (-H)}{i'}_{\rm S}
}
\label{eq:MCMC-SPU-preresult}\\
&= \frac{\sum_{m, n} p_m  p_n \sum_{k=0}^1 (-1)^k \, 2^{-N} \sum_{i}   W_{m n}^{i k} \ev{O}{\phi_{m n}^{i k}}}{\sum_{m', n'} p_{m'}  p_{n'} \sum_{k'=0}^1 (-1)^{k'} \, 2^{-N}\sum_{i'}   W_{m' n'}^{i' k'} }, 
\label{eq:MCMC-SPU-total-result}
\end{align}
which is the counterpart of Eq.~\eqref{eq:canonical-expectation-METTS} in the METTS algorithm.
From the relationship in Eq.~\eqref{eq:MCMC-SPU-total-result}, the thermal expectation value $\ev{O}_\beta$ can be obtained as a weighted average of the quantity $2^{-N}\sum_{i=0}^{2^N -1} W_{m n}^{i k} \ev{O}{\phi_{m n}^{i k}}$ with respect to sampling of pairs of orders $m$ and $n$, which is realized in Step~9 in Sec.~\ref{subsec:MCMC-SPU-procedure}.
The quantity appearing in the numerator on the right-hand side of Eq.~\eqref{eq:MCMC-SPU-total-result} can be rewritten as follows:
\begin{align}
2^{-N}\sum_{i=0}^{2^N -1} W_{m n}^{i k} \ev{O}{\phi_{m n}^{i k}}
= \qty(2^{-N} Z_{m n}^k )\ev{O}_{m n}^k ,\label{eq:MCMC-SPU_ensemble_average}
\end{align}
where $Z_{m n}^k$ on the right-hand side is provided by
\begin{align} \label{eq:MCMC-SPU-total-weight}
Z_{m n}^{k} := \sum_{i=0}^{2^N -1} W_{m n}^{i k},
\end{align}
and the corresponding quantity $\ev{O}_{m n}^k$ is defined as follows:
\begin{align} \label{eq:MCMC-SPU-each-ensemble-average}
\ev{O}_{m n}^k
:= \sum_{i=0}^{2^N -1} \frac{W_{m n}^{i k}}{Z_{m n}^k} \ev{O}{\phi_{m n}^{i k}}.
\end{align}
The weight factor $2^{-N} Z_{m n}^k = 2^{-N} \sum_{i=0}^{2^N -1} W_{m n}^{i k}$ on the right-hand side of Eq.~\eqref{eq:MCMC-SPU_ensemble_average} and the denominator in Eq.~\eqref{eq:MCMC-SPU-total-result} can be obtained as a statistical mean of $\{W_{m n}^{i k}\}$ evaluated by uniform sampling with respect to the label $i$ (see Appendix~\ref{append:reweighting} for a discussion on the efficiency of this sampling).
The value of $W_{m n}^{i k}$ represents the probability of obtaining the measurement outcome of the ancilla qubits with the result $\ket{k}_{\overline{\rm A}} \otimes \ket{0}_{\! \rm A}^{\otimes \lceil \log_2 d' \rceil}$ in Step~5 of the procedure in Sec.~\ref{subsec:MCMC-SPU-procedure}.
The quantity $\ev{O}_{m n}^k$ in Eq.~\eqref{eq:MCMC-SPU-each-ensemble-average} is calculated as the ensemble average of the observable $O$ with respect to the statistical ensemble consisting of the set of states $\{\ket{\phi_{m n}^{i k}}\}$ with the probability distribution $\{ W_{m n}^{i k}/Z_{m n}^k\}$.

As detailed below, the probability distribution $\{ W_{m n}^{i k}/Z_{m n}^k\}$ forms a stationary distribution of the Markov process $i \to j$ with respect to the label of states for the target system in the computational basis, where the Markov process is defined by Step~7 in the procedure in Sec.~\ref{subsec:MCMC-SPU-procedure}.
Starting with the probability distribution $\{W_{m n}^{i k} / Z_{m n}^k\}$, the resulting probability for label $j$ after one step of this Markov process can be derived as follows:
\begin{align}
&\sum_{i=0}^{2^N -1} \frac{W_{m n}^{i k}}{Z_{m n}^k} \qty|_{\rm S}\!\braket{j}{\phi_{m n}^{i k}}|^2 = \frac{W_{m n}^{j k}}{Z_{m n}^k}. \label{eq:MCMC-SPU-OneStep}
\end{align}
To see this, we observe that the transition probability from label $i$ to $j$ in this Markov process is expressed as follows:
\begin{align}
&\qty|_{\rm S}\!\braket{j}{\phi_{m n}^{i k}}|^2 \nonumber\\
&= \frac{1}{W_{m n}^{i k}}
\,_{\rm tot}\!\bra{i}U^\dag_{m n}\qty(\ketbra{k}_{\rm \overline{A}} \otimes \ketbra{G}_{\rm A} \!\otimes \ketbra{j}_{\rm S}) U_{m n} \ket{i}_{\rm tot} \nonumber\\
&= \frac{1}{W_{m n}^{i k}}
\,_{\rm tot}\!\bra{j}U^\dag_{m n}\qty(\ketbra{k}_{\rm \overline{A}} \otimes \ketbra{G}_{\rm A} \!\otimes \ketbra{i}_{\rm S}) U_{m n} \ket{j}_{\rm tot}. \label{eq:MCMC-SPU_transition}
\end{align}
In this transformation, we have used the Hermiticity of the block element within the representation matrix for the whole quantum circuit, which is conditioned on the measurement outcomes on the ancilla qubits, i.e.,
\begin{align}
&_{\rm \overline{A}}\bra{k}\otimes \,_{\rm {A}}\!\bra{G} \otimes \,_{\rm {S}}\!\bra{j} U_{m n} \ket{0}_{\rm \overline{A}}\otimes\ket{G}_{\rm A} \otimes\ket{i}_{\rm S} \nonumber\\
= &  \,_{\rm {S}}\!\bra{j} \qty(_{\rm \overline{A}}\bra{k}\otimes \,_{\rm {A}}\!\bra{G}  U_{m n} \ket{0}_{\rm \overline{A}}\otimes\ket{G}_{\rm A}  )^\dag \ket{i}_{\rm S} \nonumber\\
= & \,_{\rm \overline{A}}\bra{0}\otimes \,_{\rm {A}}\!\bra{G} \otimes \,_{\rm {S}}\!\bra{j} U^\dag_{m n} \ket{k}_{\rm \overline{A}}\otimes\ket{G}_{\rm A} \otimes\ket{i}_{\rm S}.
\end{align}
Based on the above observation, we can express Eq.~\eqref{eq:MCMC-SPU-OneStep} as follows:
\begin{align}
&\sum_{i=0}^{2^N -1} \frac{W_{m n}^{i k}}{Z_{m n}^k} \qty|\braket{j}{\phi_{m n}^{i k}}|^2 \nonumber\\
&= \frac{1}{Z_{m n}^k}  \,_{\rm tot}\!\bra{j}U^\dag_{m n}\qty(\ketbra{k}_{\rm \overline{A}} \otimes \ketbra{G}_{\rm A} \otimes I_{\rm S} ) U_{m n} \ket{j}_{\rm tot} \nonumber\\
&= \frac{W_{m n}^{j k}}{Z_{m n}^k} ,
\end{align}
which confirms that $\{W_{m n}^{i k} / Z_{m n}^k\}$ is a stationary distribution of the Markov process.
From this result, through a large number of cycle repetitions consisting of Steps~3--7, we can effectively evaluate Eq.~\eqref{eq:MCMC-SPU-each-ensemble-average}. This process ultimately enables the calculation of a finite-temperature expectation value, as described in Eq.~\eqref{eq:MCMC-SPU-total-result}.

\section{Implementation cost for the MCMC-SPU algorithm}\label{sec:basic-properties}

In this section, we discuss the implementation cost of the MCMC-SPU algorithm, focusing on
the number of qubits, quantum gates, and quantum circuits needed,
as well as the probability of successful postselection with respect to ancilla-qubit measurement outcomes.
As an illustrative example, we examine the one-dimensional transversal-field Ising model with the Hamiltonian:
\begin{align} \label{eq:Hamiltonian-QTFI}
H_{\rm TFI} = \sum_{j=1}^N J\sigma^x_j \sigma^x_{j+1} + h \sigma^y_j.
\end{align}
We parameterize the coefficients as $J = N^{-1} \cos^2(\theta/2)$ and $h = N^{-1}\sin^2(\theta/2)$, using a single parameter $\theta$. This parameterization arises naturally in the LCU implementation of the block encoding.

\begin{figure*}[htbp]
\includegraphics[keepaspectratio, scale=1.38]{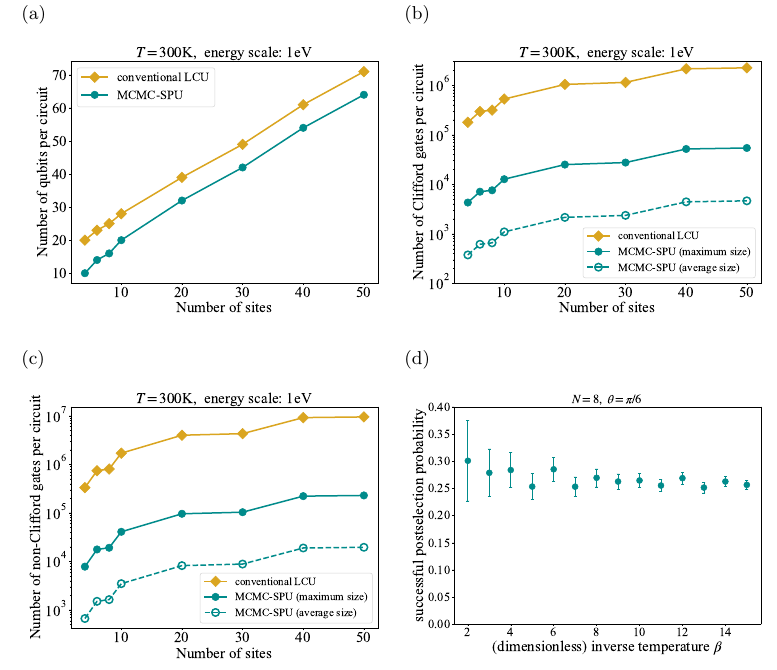}

    \caption{
    (a)--(c)
    Resources required for implementing the proposed MCMC-SPU algorithm and that for the conventional form of the LCU.
    The Hamiltonian is given by the transverse-field Ising model in Eq.~\eqref{eq:Hamiltonian-QTFI}. The maximum absolute value of the eigenenergies is set to be $1~{\rm eV}$, and the absolute temperature is set to be $T=300{\rm K}$. 
    The truncation order $d$ is determined as $d=82$ based on Eq.~\eqref{eq:required-order-LCU} with $\nu = 0.1$.
    (a) Total number of qubits including ancilla qubits for each quantum circuit.
    \textcolor{black}{Dark cyan} circles (brown diamonds) represent the total number of qubits required for implementing the proposed MCMC-SPU algorithm (the conventional LCU method as a part of the QMETTS algorithm).
    (b)--(c) Gate count of each quantum circuit, where we decompose each quantum circuit into a set of Clifford gates and single-qubit non-Clifford rotation gates.
    (b) Number of Clifford gates appearing in each quantum circuit.
    Brown diamonds represent the number of Clifford gates 
    required for implementing the conventional LCU method as a part of the QMETTS algorithm.
    \textcolor{black}{Dark cyan} filled (empty) circles represent the maximum number (the average weighted by the sampling probability)  of Clifford gates per quantum circuit appearing in stochastically sampling pairs $(m, n)$ of orders in the proposed MCMC-SPU algorithm (Steps~1.--2. in Sec.~\ref{subsec:MCMC-SPU-procedure}).
    (c) Number of non-Clifford single-qubit rotation gates appearing in each quantum circuit.
    Brown diamonds represent the number of non-Clifford single-qubit rotation gates 
    required for implementing the conventional LCU method as a part of the QMETTS algorithm.
    \textcolor{black}{Dark cyan} filled (empty) circles represent the maximum number (the average weighted by the sampling probability)  of those gates per quantum circuit in stochastically sampling pairs $(m, n)$ of orders in the proposed MCMC-SPU algorithm.
    (d) Successful postselection probability for the quantum circuits in the proposed MCMC-SPU algorithm. The plot shows the unweighted average values with error bars representing the standard errors calculated over all the possible pairs $(m, n)$ of orders satisfying $m\le d$ and $n\le d$, where $d$ is derived from Eq.~\eqref{eq:required-order-LCU} for each value of the dimensionless inverse temperature $\beta$.
    }
    \label{fig:resource_implementation} 
\end{figure*}

\subsection{Number of qubits and quantum gates}\label{subsec:resource-implementation}

Initially, we discuss the number of qubits and quantum gates required for implementing each quantum circuit in the MCMC-SPU algorithm. The resources required are influenced by the truncation order $d$ of the LCU method [see Eq.~\eqref{eq:expansion-QITE}], which is determined by the inverse temperature $\beta$ for the simulation and precision $\nu$ of the approximation. Specifically, to ensure that the $d$-th order approximation $\tilde{f}(\cdot)$ of $f(\cdot)$ satisfies Eq.~\eqref{eq:error-fA},
the required truncation order of LCU is provided by the following~\cite{Tosta2023,Sachdeva2013}:
\begin{align} \label{eq:required-order-LCU}
d = \sqrt{2 \log(\frac{4e^{\beta/2}}{\nu}) \cdot \max\!\qty[\frac{e^2 \beta}{2}, \, \log(\frac{2e^{\beta/2}}{\nu})]}
\end{align}
\textcolor{black}{exhibiting} a nearly linear scaling with $\beta$ [see also Eq.~\eqref{eq:error-observable}].
By contrast, when stochastically sampling order pairs $(m, n)$ with $m\le d$ and $n \le d$ in the MCMC-SPU algorithm (Step~1 in Sec.~\ref{subsec:MCMC-SPU-procedure}), the weighted average $d_{\rm average}$ of a sampled value of order $n$ is evaluated as follows~\cite{Tosta2023}:
\begin{align}
\label{eq:average-order-LCU}
&d_{\rm average}
= \frac{1}{||c||_1} \sum_{n=1}^d n \, c_n
\simeq e^{-\beta/2} \sum_{n=1}^d 2n \, I_n(\beta/2) \nonumber\\
&\le \frac{\beta}{2} e^{-\beta/2} \, \qty[I_0(\beta/2) + I_1(\beta/2) - I_d (\beta/2) - I_{d+1} (\beta/2) ],
\end{align}
the most right-hand side of which exhibits a scaling $\mathcal{O}(\sqrt{\beta})$ as $\beta \to \infty$.
In this transformation, we have used the relationships $n \, I_n(\beta/2) = (\beta/4) \qty[I_{n-1}(\beta/2) - I_{n+1}(\beta/2)  ]$ and
\begin{align}
    \qty| ||c||_1 - e^{\beta/2} | \le \nu
    \label{eq:bound_c_norm}
\end{align}
with a sufficiently small value of $\nu$~\cite{Tosta2023}.
The differing scaling behaviors between $d$ and $d_{\rm average}$ as $\beta\to \infty$ highlight a significant advantage of stochastic sampling, significantly reducing implementation resources~\cite{Tosta2023} in our MCMC-SPU algorithm.
However, this algorithm incorporates sampling overhead as discussed in Sec.~\ref{subsec:number-circuits}.

The inverse temperature $\beta$ determines the resource required for implementing each quantum circuit, based on sampled polynomial orders in the series expansion of Eq.~\eqref{eq:expansion-QITE} with the truncation order $d$, given by Eq.~\eqref{eq:required-order-LCU}. Each polynomial employs $\qty(W_{ H})^k$ via Eq.~\eqref{eq:modified_Bessel_block_encoding}, with $W_{ H}$ defined in Eq.~\eqref{eq:qubitization-walk-operator} alongside Eq.~\eqref{eq:reflection}, determining the number of qubits and gate counts for each quantum circuit. The specific form of $W_{ H}$ depends on the Hamiltonian.
In our implementation, ${\rm PREP}_H$ in Eq.~\eqref{eq:reflection} for an $N$-site model is represented as $H^{\otimes \lceil \log_2 N \rceil}\otimes R_Z(\theta)$ with Hadamard gates $H$ and a single-qubit rotation gate $R_Z(\theta)$.
Furthermore, the operator $2 \ket{0}_{{\rm A}, H} \,_{{\rm A}, H}\bra{0}  - I_{\rm A} $ is implemented using a multicontrolled phase gate, and ${\rm SELECT}_H$ in Eq.~\eqref{eq:qubitization-walk-operator} takes the form as shown in Fig.~\ref{fig:coherentLCU}(b). 
In implementing multicontrolled unitary gates $C^{N_{\rm c}}(U)$ for these operators, we utilize a usual nesting structure by recursively constructing a $C^{N_{\rm c}-1}(U)$ gate placed between two Toffoli gates~\cite{Barenco1995, Nielsen2000},
each of which is assumed to be implemented with three single-qubit rotation gates~\cite{Nielsen2000}.
Figures~\ref{fig:resource_implementation}(a)--(c) show the number of qubits and quantum gates required for implementing each quantum circuit, where the Hamiltonian is given by the transverse-field Ising model in Eq.~\eqref{eq:Hamiltonian-QTFI}.
The results are independent of value of $\theta$.
The eigenvalues of the Hamiltonian are rescaled so that the maximum absolute value of the eigenenergies is set to be $1~{\rm eV}$, and the absolute temperature is set to $T=300~{\rm K}$, determining the LCU truncation order $d$ as $d=82$ as per Eq.~\eqref{eq:required-order-LCU} with $\nu = 0.1$.

Figure~\ref{fig:resource_implementation}(a) presents the total number of qubits including ancilla qubits for each quantum circuit. 
Both the conventional LCU implementation (brown diamonds) and the proposed MCMC-SPU (\textcolor{black}{dark cyan} circles) show nearly linear qubit scaling with the number $N$ of sites, while the latter requires fewer ancilla qubits.

We next discuss the gate count of each quantum circuit. We decompose each quantum circuit into a set of Clifford gates and single-qubit non-Clifford rotation gates. It is worth noting that we do not further decompose single-qubit non-Clifford rotation gates into a large number of Clifford gates and T gates, based on the assumption that such decomposition is beyond the capability of early-FTQC devices (see Sec.~\ref{sec:discussion} for a discussion on direct implementation of single-qubit rotation gates). Figure~\ref{fig:resource_implementation}(b) shows the number of Clifford gates appearing in each quantum circuit. 
In our MCMC-SPU algorithm, the maximum number of Clifford gates per circuit  (\textcolor{black}{dark cyan} filled circles) is about two orders of magnitude smaller than that in the conventional LCU implementation (brown diamonds).
Furthermore, the average number of Clifford gates per circuit weighted by the sampling probability of each pair order in our MCMC-SPU (\textcolor{black}{dark cyan} empty circles) is about three orders of magnitude smaller than that in the conventional implementation (brown diamonds).

Figure~\ref{fig:resource_implementation}(c) demonstrates the number of non-Clifford single-qubit rotation gates appearing in each quantum circuit.
Similarly to Clifford gates, the maximum number of these gates per circuit in our MCMC-SPU's stochastic sampling (\textcolor{black}{dark cyan} filled circles) is about two orders of magnitude smaller compared to the conventional LCU implementation (brown diamonds).
Moreover, the average number of these gates per circuit weighted by the sampling probability of each order pair in our MCMC-SPU (\textcolor{black}{dark cyan} empty circles) is about three orders of magnitude smaller than that in the conventional implementation (brown diamonds).

In summary, our MCMC-SPU algorithm requires fewer qubits and quantum gates compared to the conventional LCU implementation. In particular, the required number of quantum gates is reduced by orders of magnitude.
Even if the absolute temperature is set to be different from $T=300~{\rm K}$, the scaling of required qubits and quantum gates in relation to the number of sites remains unchanged.
Furthermore, the MCMC-SPU offers a better scaling for implementation resources with increasing $\beta$, making it suitable for the early-FTQC era. This is because the average value $d_{\rm average}$ of a sampled order scales as $\mathcal{O}(\sqrt{\beta})$ in the MCMC-SPU, while the conventional LCU implementation deterministically involves a circuit depth $\mathcal{O}\qty(\sum_{k=0}^dk)=\mathcal{O}(\beta^2)$ and requires $\mathcal{O}\qty(\log_2 d)=\mathcal{O}\qty(\log_2 \beta)$ ancilla qubits for superposition [see Eq.~\eqref{eq:required-order-LCU}].

\subsection{Probability of successful postselection}\label{subsec:probability-PS}

We discuss the probability of successful postselection on the measurement outcomes of ancilla qubits labeled $\rm A$ and $\overline{\rm A}$ for each quantum circuit (see Fig.~\ref{fig:circuit-MCMC-SPU} and Step~5 in Sec.~\ref{subsec:MCMC-SPU-procedure}) appearing in the MCMC-SPU algorithm. 
Using the Qulacs package~\cite{Suzuki2021}, we perform a numerical simulation on the one-dimensional transverse-field Ising model in Eq.~\eqref{eq:Hamiltonian-QTFI}  with $N=8$ and $\theta = \pi/6$ as an illustrative example.
Figure~\ref{fig:resource_implementation}(d) presents the numerical results of the successful postselection probability for the quantum circuits in our MCMC-SPU algorithm. The plot displays the unweighted average value across all possible pairs $(m, n)$ of orders satisfying $m\le d$ and $n\le d$ (see Appendix~\ref{apppend:detail_PS_prob} for the detailed data), where $d$ is derived from Eq.~\eqref{eq:required-order-LCU} for each value of the dimensionless inverse temperature $\beta$.
The results presented in Fig.~\ref{fig:resource_implementation}(d) suggests a nearly constant successful postselection probability, avoiding exponential decay with $\beta$.
This fact stands in stark contrast to an exponential decay with $\beta$ observed in the conventional LCU implementations (see Sec.~\ref{subsubsec:PSprob-QITE}).
It is important to note that the lower bound on the successful postselection probability for each quantum circuit in implementing the single-ancilla LCU for a general matrix function is obtained in Ref.~\cite{Childs2012}.
In our implementation, this lower bound for each quantum circuit is independent of value $\beta$ as its influence is confined to the probability distribution for stochastic sampling of orders through the coefficients $\{c_n\}$ in Eq.~\eqref{eq:MCMC-SPU-coefficient}.
Detailed analysis on the $\beta$ dependence of the weighted average of the lower bounds over all the possible order pairs is left for future research.

\subsection{Number of quantum circuits}\label{subsec:number-circuits}

We now examine the number of quantum circuits required for our MCMC-SPU algorithm.
Since the numerator (denominator) on the right-hand side of Eq.~\eqref{eq:MCMC-SPU-preresult} is a statistical mean of a stochastic variable that lies in the range $\qty[- ||O||_{\rm op} \cdot ||c||_1^2, \, ||O||_{\rm op} \cdot ||c||_1^2]$ ($\qty[-||c||_1^2, \, ||c||_1^2]$), the Hoeffding's inequality~\cite{Hoeffding1963} tells us that the required number $N_{\epsilon, \delta}$ of order pairs, equivalent to the number of quantum circuits, is specified as follows:
\begin{align} \label{eq:number_Qcircuit}
N_{\epsilon, \delta} = \mathcal{O} \qty[ \frac{ ||c||_1^4}{\epsilon^2} \log(\frac{2}{\delta}) ]
=\mathcal{O} \qty[ \frac{ e^{2\beta} }{\epsilon^2} \log(\frac{2}{\delta}) ]
\end{align}
to ensure that the error of the statistical mean from the true expectation value is bounded by $\epsilon$ with probability $1 -\delta$.
This analysis focuses solely on the statistical error $\epsilon$, excluding the sampling overhead associated with calculating the quantum mechanical expectation value.
In Eq.~\eqref{eq:number_Qcircuit}, we use Eq.~\eqref{eq:bound_c_norm} for the $d$-th-order approximation $\tilde{f}(\cdot)$ of $f(\cdot)$ satisfying Eq.~\eqref{eq:error-fA}~\cite{Tosta2023} while omitting the scaling of a sufficiently small quantity $\nu$.
The scaling of $N_{\epsilon, \delta}\propto e^{2\beta}$ for the number of circuits in our MCMC-SPU algorithm aligns with the required number $N_{\rm PS}$ of trials until realizing a successful postselection in conventional LCU implementations (see Eq.~\eqref{eq:PSprob_generalLCU} and the following argument).
This fact, together with the results in Sec.~\ref{subsec:probability-PS},
indicates that the required total number of quantum-circuit executions, defined as the product of the number of circuits and the inverse of the successful postselection probability, exhibits a scaling $\mathcal{O}\qty(e^{2\beta})$ similar to that of the conventional LCU implementation (see also Ref.~\cite{Tosta2023}).
Note that our MCMC-SPU algorithm allows for parallel executions of various quantum circuits $U_{m n}$'s on independent quantum devices in our MCMC-SPU algorithm unlike the conventional implementation (see also Fig.~\ref{fig:overview-MCMCSPU}).


\section{Numerical demonstration of the MCMC-SPU algorithm}\label{sec:numerical-demonstration}

\begin{figure*}[htbp]
\includegraphics[keepaspectratio, scale=1.38]{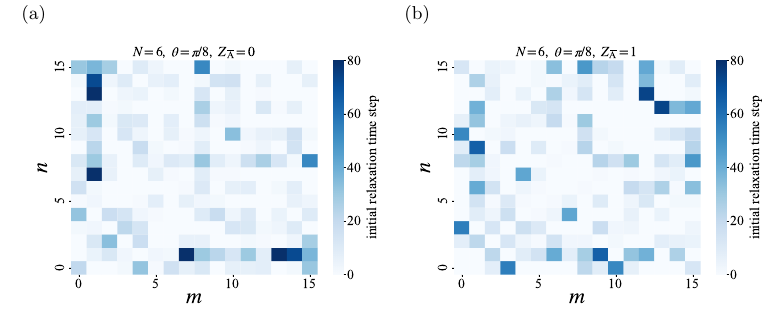}
    \caption{ 
    initial relaxation time step for each quantum circuit appearing in the MCMC-SPU algorithm for the one-dimensional transverse-field Ising model with six sites.
    The plots show the initial relaxation time step evaluated conditioned on the measurement outcome of the ancilla qubit for superposition with label $\overline{\rm A}$ being (a) $Z_{\overline{\rm A}} = 0$ and  (b) $Z_{\overline{\rm A}} = 1$ for the quantum circuit $U_{m n}$. This quantum circuit involves unitary gates $U_m$ and $U_n$ corresponding to a pair of orders $m$ and $n$ stochastically sampled from the expansion of an imaginary-time evolution.
    In Fig.~(b), although the results are plotted as zero for $m=n$, this representation is solely for convenience. It actually indicates that no MCMC sampling is conducted for these specific index combinations (see Eq.~\eqref{eq:W_no_contribution} and the following argument).
    }
    \label{fig:initial_relaxation}
\end{figure*}

We present a numerical demonstration of the proposed MCMC-SPU algorithm by applying it to the one-dimensional transverse-field Ising model in Eq.~\eqref{eq:Hamiltonian-QTFI} with $N=6$ and $\theta = \pi/8$ as an illustrative example. The maximum absolute value of the eigenenergies is set to be $0.1~{\rm eV}$.
The truncation order $d$ is determined using Eq.~\eqref{eq:required-order-LCU} with $\nu=0.002$, 
and this demonstration does not involve simulated noise.
As is common in MCMC simulations~\cite{Geyer1992, Gilks1995, Cowles1996, Brooks1998}, we first evaluate the initial relaxation time step $n_{\rm relax}$, ensuring that the influence of a randomly chosen initial state diminishes over time.
To minimize this effect, we focus on a statistical ensemble consisting of samples after the initial relaxation.
To appropriately evaluate the number $n_{\rm eff}$ of effectively independent samples, we examine the autocorrelation time step $\tau$ after the initial relaxation period. Specifically, $n_{\rm eff}$ is evaluated as the number of samples $n_{\rm sample}$ belonging to the statistical ensemble divided by twice the integrated autocorrelation time step $\tau$, i.e., $n_{\rm eff} = n_{\rm sample}/(2\tau)$.
Finally, we demonstrate that our MCMC-SPU algorithm accurately computes the finite-temperature value of an observable.
In the following subsections, we discuss each step of the above procedure.

\subsection{Initial relaxation}\label{subsec:initical-relaxation}

We first consider the initial relaxation time step $n_{\rm relax}$, during which the dependence on the randomly chosen initial state gradually vanishes.
The first $n_{\rm relax}$ steps are called the burn-in period, and samples from this period are excluded from the statistical ensemble.

In our analysis, for each quantum circuit appearing in the MCMC-SPU algorithm (see Fig.~\ref{fig:circuit-MCMC-SPU}), we focus on two qualitatively different initial states: $\ket{\psi_{\rm F}}=\ket{000000}$ and $\ket{\psi_{\rm AF}}=\ket{010101}$. We then generate a series of measurement outcomes of an observable through MCMC sampling for each initial state.
The initial relaxation time step is estimated as the time step at which the two series of results generated by the two initial states $\ket{\psi_{\rm F}}$ and $\ket{\psi_{\rm AF}}$ become statistically indistinguishable (see Appendix~\ref{append:Gelman-Rubin} for technical details of the analysis on the initial relaxation).

Figure~\ref{fig:initial_relaxation} presents the numerical results of the initial relaxation time step for each quantum circuit $U_{m n}$ in the MCMC-SPU algorithm applied to the transverse-field Ising model. Figure~\ref{fig:initial_relaxation}(a) [(b)] shows the initial relaxation time step evaluated conditioned on the ancilla-qubit measurement outcome for superposition, labeled $\overline{\rm A}$, being zero [one] for the quantum circuit $U_{m n}$, which involves unitary gates $U_m$ and $U_n$ corresponding to orders $m$ and $n$ stochastically sampled from the expansion of an imaginary-time evolution (Steps~1--2 in Sec.~\ref{subsec:MCMC-SPU-procedure}).
For most order pairs illustrated in Fig.~\ref{fig:initial_relaxation}, which are relevant to the following numerical simulation, the resulting values of the initial relaxation time step are less than 40 (see Appendix~\ref{append:scaling_overhead} for the system-size dependence). These results are comparable to that of the conventional LCU implementation (see Sec.~\ref{subsubsec:LCU-QITE}), which was numerically evaluated as $10$.

\begin{figure*}[htbp]\includegraphics[keepaspectratio, scale=1.38]{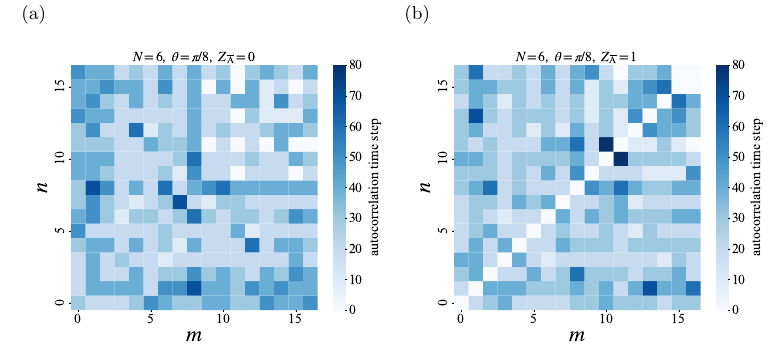}
    \caption{
    Autocorrelation time step for each quantum circuit in the MCMC-SPU algorithm for the one-dimensional transverse-field Ising model with six sites.
    The plots show the autocorrelation time step evaluated conditioned on the measurement outcome of the ancilla qubit for superposition with label $\overline{\rm A}$ being (a) $Z_{\overline{\rm A}} = 0$ and  (b) $Z_{\overline{\rm A}} = 1$ for quantum circuit $U_{m n}$. This quantum circuit involves unitary gates $U_m$ and $U_n$ corresponding to a pair of orders $m$ and $n$ stochastically sampled from the expansion of an imaginary-time evolution.
    In Fig.~(b), although the results are plotted as zero for $m=n$, this representation is solely for convenience. It actually indicates that no MCMC sampling is conducted for these specific index combinations (see Eq.~\eqref{eq:W_no_contribution} and the following argument).
    }
    \label{fig:autocorrelation}
\end{figure*}

\subsection{Autocorrelation}\label{subsec:autocorrelation}

In our MCMC-SPU algorithm, we assess the autocorrelation time step following the initial relaxation for each quantum circuit.
This time step represents
the number of consecutive samples in the MCMC sampling where a statistical correlation remains significant. The effective number $n_{\rm eff}$ of independent samples is determined by dividing the number $n_{\rm sample}$ of all the samples belonging to the statistical ensemble by twice the integrated autocorrelation time step $\tau$, i.e., $n_{\rm eff} = n_{\rm sample}/(2\tau)$.
We employ the jackknife method~\cite{Quenouille1949_jackknife, Quenouille1956, Tukey1958, Efron1994} to determine the autocorrelation time step numerically  (see Appendix~\ref{appned:jackknife-autocorrelation} for details of the jackknife method).

Figure~\ref{fig:autocorrelation} shows the numerical results of the autocorrelation time step for each quantum circuit $U_{m n}$  (see Fig.~\ref{fig:circuit-MCMC-SPU}) in the proposed MCMC-SPU algorithm. Figure~\ref{fig:autocorrelation}(a) [(b)] shows the results evaluated under the condition that the measurement outcome of the ancilla qubit for superposition, labeled $\overline{\rm A}$, is $Z_{\overline{\rm A}}=0$ [$Z_{\overline{\rm A}}=1$] for the quantum circuit $U_{m n}$.
For most order pairs shown in Fig.~\ref{fig:autocorrelation}, the numerical results are less than 60 (see Appendix~\ref{append:scaling_overhead} for the system-size dependence). These results are comparable to that of the conventional LCU implementation (see Sec.~\ref{subsubsec:LCU-QITE}), which was numerically evaluated to be $10$.
Together with the findings in Sec.~\ref{subsec:initical-relaxation}, these results suggest that the statistical sampling overhead for each quantum circuit in the MCMC-SPU algorithm is not considerably large compared with that in the QMETTS with the conventional LCU implementation.

\subsection{Thermal average of an observable}\label{subsec:thermal-average}

\begin{figure}[htbp]
    \includegraphics[keepaspectratio, scale=0.30]{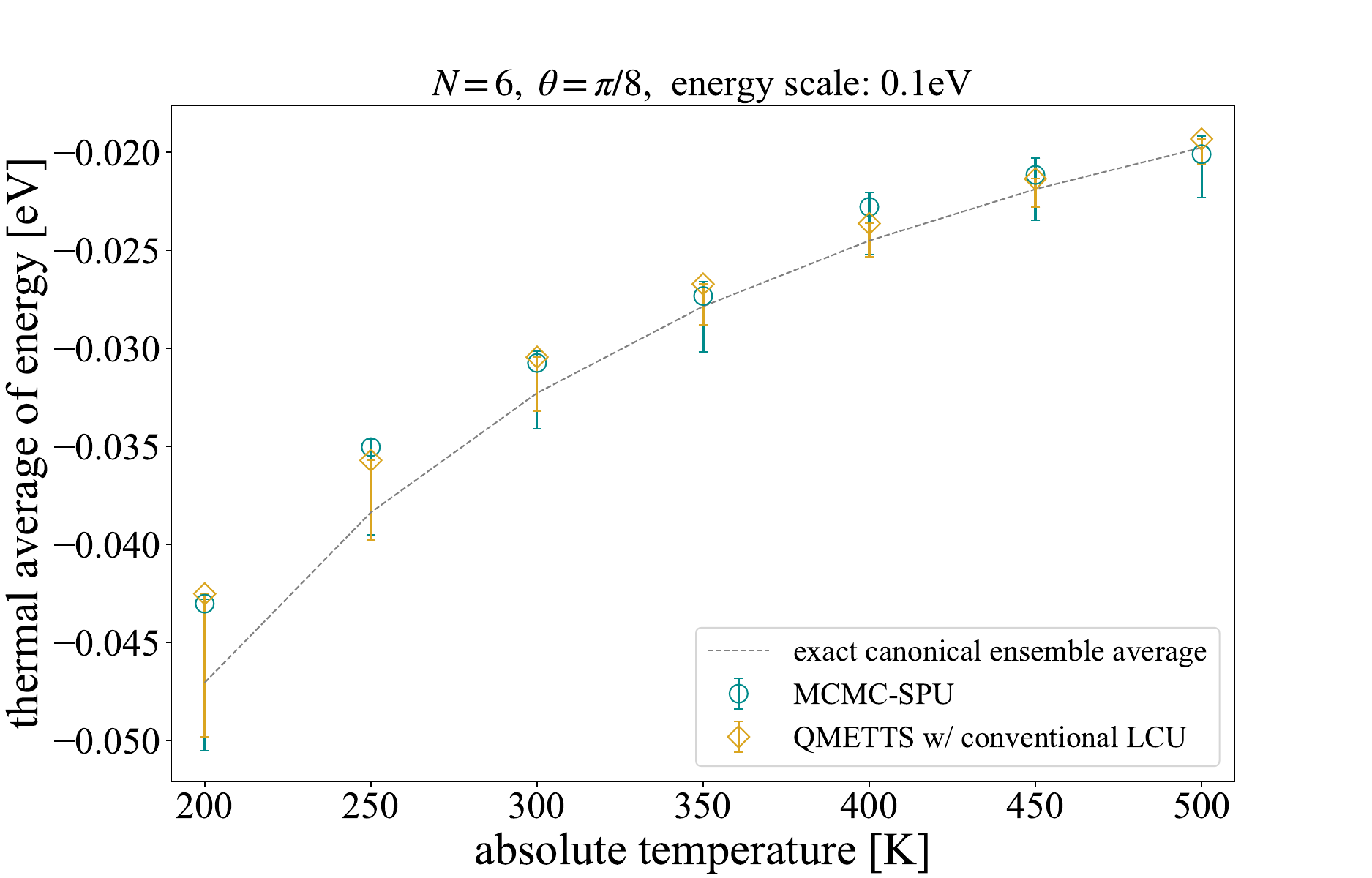}
    \caption{
    Thermal-equilibrium expectation value of the energy calculated with the MCMC-SPU algorithm (\textcolor{black}{dark cyan} circles) and with the QMETTS with the conventional LCU implementation (brown diamonds).
    The error bars include the statistical error and the truncation error [see Eq.~\eqref{eq:error-observable}] from the exact canonical ensemble average (gray dashed curve).
    The statistical error is calculated based on the standard errors for each numerator and denominator on the right-hand side of Eq.~\eqref{eq:MCMC-SPU-total-result}.
    The contribution of the truncation error is included only below the plotted points.
    The Hamiltonian is given by the one-dimensional transverse-field Ising model with parameters $N=6$ and $\theta=\pi/8$ [see Eq.~\eqref{eq:Hamiltonian-QTFI}],
    where the maximum absolute value of the eigenenergies is set to be $0.1~{\rm eV}$.
    In the numerical simulation, $\epsilon$ in Eq.~\eqref{eq:number_Qcircuit} is set to be  $\epsilon=0.1$.
    In the calculation with the MCMC-SPU (QMETTS) algorithm, the burn-in period and the autocorrelation time step are set to be 100 (10) based on the results in Secs.~\ref{subsec:initical-relaxation} and \ref{subsec:autocorrelation}.
    }
\label{fig:thermal_energy}
\end{figure}

Based on the analysis from previous subsections, we calculate the thermal-equilibrium expectation value of an observable at finite temperatures using our MCMC-SPU algorithm.
Figure~\ref{fig:thermal_energy} shows the thermal expectation value of the energy calculated with our MCMC-SPU algorithm (\textcolor{black}{dark cyan} circles) and the QMETTS using the conventional LCU implementation (brown diamonds). In this figure, error bars reflect both the statistical error and the uniform deviation from the exact canonical ensemble average (gray dashed curve) owing to the truncation in the polynomial expansion of an imaginary-time evolution [see Eq.~\eqref{eq:error-observable}].
For the MCMC-SPU, the statistical error is derived from the standard errors of the numerator and the denominator on the right-hand side of Eq.~\eqref{eq:MCMC-SPU-total-result}.
The truncation error only contributes to below the plotted points.
Both algorithms yield results that closely match the exact canonical ensemble average (gray dashed curve) within the error bars.

\section{Discussion and Conclusion}\label{sec:discussion}

In this paper, we have introduced the MCMC-SPU algorithm, designed to compute the thermal-equilibrium expectation value of an observable at finite temperatures. Our proposal is well-suited for implementation on early-FTQC devices.
The implementation involves a set of simple quantum circuits, each featuring a pair of unitary gates $U_m$ and $U_n$ (see Fig.~\ref{fig:circuit-MCMC-SPU}) stochastically sampled from the set of the unitary gates $\{U_0, U_1, \cdots, U_d\}$ used in the conventional LCU method (see Fig.~\ref{fig:coherentLCU}). For each quantum circuit consisting of these two unitary gates, a statistical ensemble unique to the circuit is generated as the stationary distribution of the MCMC sampling.

We have demonstrated that the stochastic sampling of polynomial orders in the MCMC-SPU algorithm significantly reduces the need for implementation resources like qubits and quantum gates compared to the conventional LCU implementation. The number of required quantum gates is reduced by orders of magnitude (see Sec.~\ref{subsec:resource-implementation}), enhancing our algorithm's suitability for the early-FTQC era (see the discussion below about the possibility of implementation on a prototypical device).
We have also discussed that the total number of quantum-circuit executions needed, defined as the product of the number of circuits and the inverse of the successful postselection probability, scales as $\mathcal{O}\qty(e^{2\beta})$, similarly to the conventional LCU implementation (see Secs.~\ref{subsec:probability-PS} and \ref{subsec:number-circuits}). \textcolor{black}{The classical computational overhead, such as quantum-circuit sampling, and its scaling are summarized in Appendix~\ref{append:classical-overhead}.}

To validate our proposal, we have performed a numerical simulation of the MCMC-SPU algorithm, applying it to the one-dimensional transverse-field Ising model as a case study.
We have found that the statistical sampling overhead, which includes initial relaxation and autocorrelation, is not significantly larger for the MCMC-SPU algorithm compared to the QMETTS with the conventional LCU implementation (see Secs.~\ref{subsec:initical-relaxation} and \ref{subsec:autocorrelation}).
Furthermore, we have calculated the thermal-equilibrium expectation value of the energy at finite temperatures. The results of the MCMC-SPU algorithm closely align with the exact values, staying within the bounds of statistical and truncation errors (see Sec.~\ref{subsec:thermal-average}).

We now discuss the possibility of implementing our MCMC-SPU algorithm on a prototypical example of early-FTQC devices, \textcolor{black}{taking account of quantum hardware noise. Specifically, we focus} on the Space-Time efficient Analog Rotation quantum computing architecture (STAR architecture), which was recently proposed in Ref.~\cite{Akahoshi2024}. 
This architecture enables universal quantum computation by combining fault-tolerant Clifford gates with non-fault-tolerant analog rotation gates, eliminating the need for T-gate decomposition. \textcolor{black}{In this architecture, the} analog rotation gates, implemented through gate teleportation with specific magic states, are the primary source of logical errors, \textcolor{black}{which are induced by quantum hardware noise, characterized by the physical error rate $p_{\rm phys}$}. The corresponding logical error rate $P_{\rm rotation}$ is evaluated as $P_{\rm rotation}= 2p_{\rm phys}/15 + \mathcal{O}\qty(p_{\rm phys}^2)$\textcolor{black}{~\cite{Akahoshi2024}}.
To minimize these errors, probabilistic error cancellation can be used with a reasonable sampling overhead, provided the number of analog rotation gates $N_{\rm rotation}$ satisfies $\qty( 2 N_{\rm rotation} ) P_{\rm rotation} \le 2$~\cite{Endo2018PRX}.
Here, $2 N_{\rm rotation}$ represents the effective number of analog rotation gates, where the factor $2$ originates from the average number of trial repetitions until the successful gate teleportation~\cite{Akahoshi2024}.
The condition for the successful error suppression can be rewritten as $p_{\rm phys} \le (15/2) N_{\rm rotation}^{-1}$.
Referring to the data in Fig.~\ref{fig:resource_implementation}(c), we assess the physical error rate required to implement the MCMC-SPU and the QMETTS algorithms with the conventional LCU implementation. For models with approximately fifty sites, which are challenging for classical simulations, the conventional LCU implementation demands a physical error rate of around $p_{\rm phys} \simeq 10^{-6}$, which is currently beyond the reach of existing technology. 
Conversely, our MCMC-SPU algorithm requires a physical error rate $p_{\rm phys} \simeq \mathcal{O} \qty( 10^{-4} )$ for the maximum-sized quantum circuits and $p_{\rm phys} \simeq \mathcal{O} \qty( 10^{-3} )$ for average-sized circuits. The latter is within the scope of current technology, depending on the hardware utilized~\cite{Harty2014, Srinivas2021, Clark2021, Moses2023, daSilva2024}.
Even for more complicated models like fermionic systems, the gate counts and required physical error rates are expected to scale similarly, under general assumptions such as spatial translational invariance and locality of interactions.

We discuss some remarks on our method compared with other possible techniques for simulating thermal-equilibrium systems at finite temperatures.
First, the amplitude-amplification technique~\cite{Poulin2009, Chowdhury2017, vanApeldoorn2020, Silva2023, Brassard2002, Berry2014, Berry2015, Berry2015PRL, Low2019muliproduct} can be integrated before measuring and postselecting on ancilla qubits in the conventional LCU method, improving the successful postselection probability~\cite{Silva2023,Tosta2023}. A similar technique could also be applied to the MCMC-SPU algorithm, although the MCMC-SPU does not significantly suffer from decay in postselection probability (see Sec.~\ref{subsec:probability-PS} and Appendix~\ref{apppend:detail_PS_prob}). Despite its potential usefulness, inserting this step leads to a significant increase in circuit depth.
%
%
Second, while our method involves generating statistical ensembles to simulate finite-temperature systems, similarly to the METTS algorithm, we could alternatively employ an approach based on pure states such as the thermal pure quantum (TPQ) states~\cite{Sugiura2012,Sugiura2013, Hyuga2014, Endo2018, Iwaki2021, Tsutsui2022}. This approach avoids explicitly generating the thermal statistical ensemble, but creating a TPQ state involves overhead in preparing a Haar-random state or a unitary $t$-design~\cite{Powers2022, Coopmans2023} as an initial state.
The possibilities raised above are beyond the scope of this paper and are left for interesting future problems.

A possible application of our technique is simulating various physical quantities in more complicated quantum many-body systems, such as electronic conductivity, magnetic susceptibility, and the Green's functions in strongly correlated electron systems. This could potentially contribute to studying real materials, which we leave for interesting future work.
We hope that this work stimulates further investigation into quantum algorithms that offer advantages in simulating large-scale quantum many-body systems.

\section*{Acknowledgement}
The authors are grateful to Riki~Toshio and Yutaro~Akahoshi for helpful comments and discussions. 
N.M. acknowledges Masatoshi~Ishii for technical support in performing numerical simulations.

\appendix


\section{Efficiency of sampling the weight factor}\label{append:reweighting}

\begin{figure}[tbp]
    \includegraphics[keepaspectratio, scale=0.23]{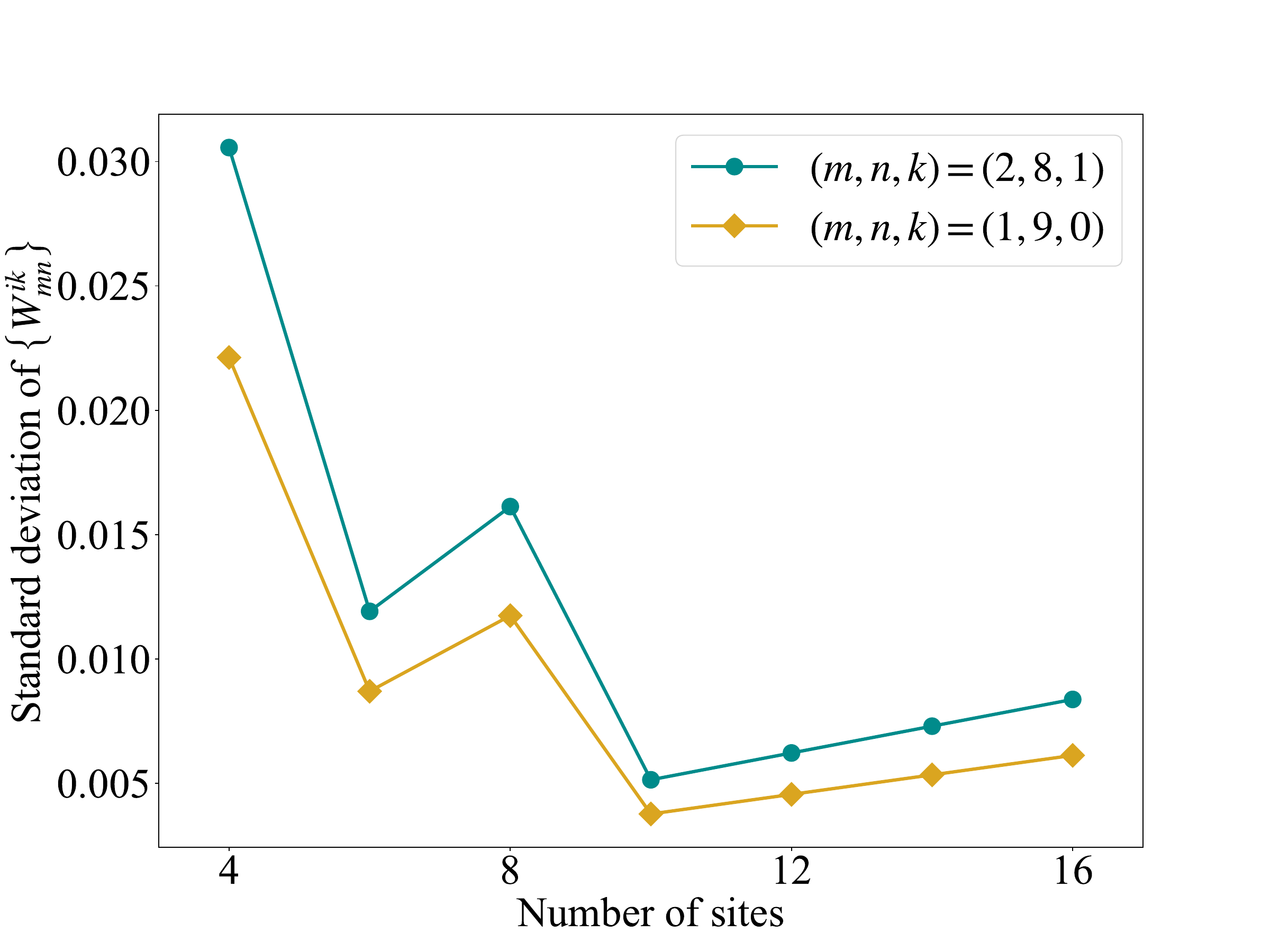}
    \caption{
    System-size dependence of the standard deviation of the set of values of term $W_{m n}^{i k}$. The \textcolor{black}{dark cyan} circles [brown diamonds] represent the results for values sampled uniformly with respect to index $i$ while $(m, n, k)$ is fixed to be $(2,8,1)$ [$(1,9,0)$].
    }
\label{fig:stdev_weight}
\end{figure}

In this appendix, we discuss the efficiency of sampling the quantity $W_{m n}^{i k}$, crucial for evaluating the numerator and the denominator on the right-hand side of Eq.~\eqref{eq:MCMC-SPU-total-result} (see also Eqs.~\eqref{eq:MCMC-SPU-total-weight} and \eqref{eq:MCMC-SPU-each-ensemble-average} and the following discussion in Sec.~\ref{subsec:MCMC-SPU-theory}).
The term $2^{-N} Z_{m n}^k = 2^{-N} \sum_{i=0}^{2^N -1} W_{m n}^{i k}$, which appears in both the numerator and the denominator on the right-hand side of Eq.~\eqref{eq:MCMC-SPU-total-result}, can be obtained as a statistical mean of $\{W_{m n}^{i k}\}$ evaluated by uniform sampling with respect to label $i$. The sampling efficiency relies on how closely the values of $\{W_{m n}^{i k}\}$ for a fixed combination of $(m, n, k)$ resemble a uniform distribution.
Figure~\ref{fig:stdev_weight} displays the standard deviation of the $W_{m n}^{i k}$ as a measure of this uniformity for certain representative parameter combinations $(m, n, k)$, using the same model and conditions as in Sec.~\ref{sec:numerical-demonstration}. The \textcolor{black}{dark cyan} circles [brown diamonds] represent results for values sampled uniformly with respect to the index $i$ while $(m, n, k)$ is fixed to be $(2,8,1)$ [$(1,9,0)$].
These results indicate that the standard deviation does not significantly increase with system size, implying that the distribution of $W_{m n}^{i k}$ remains nearly uniform within our simulation range.
This result supports the efficiency of our strategy to evaluate $Z_{m n}^k$ by sampling label $i$ uniformly.

\section{System-size dependence of the statistical sampling overhead}\label{append:scaling_overhead}

\begin{figure}[tbp]
    \includegraphics[keepaspectratio, scale=0.235]{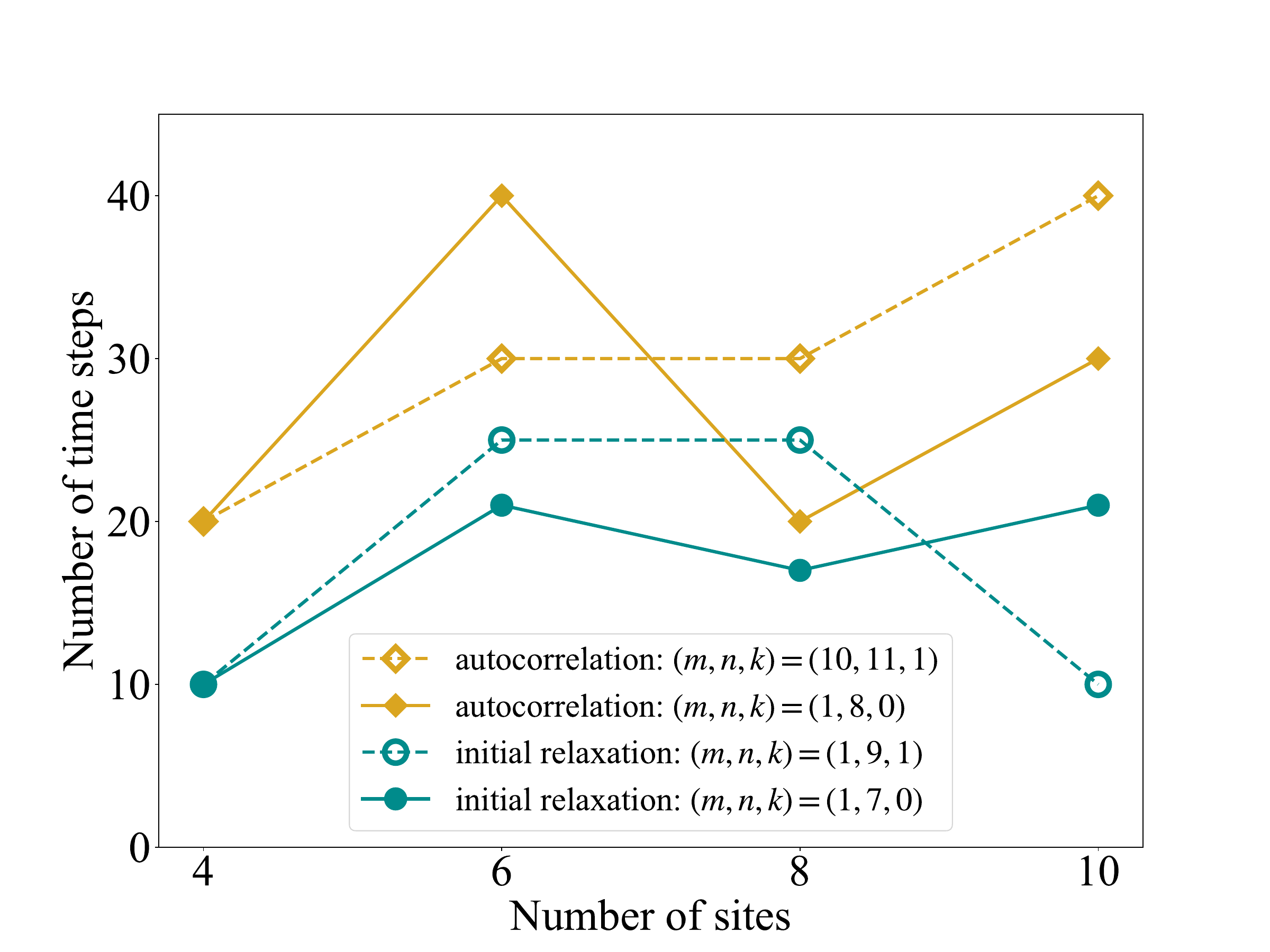}
    \caption{
    System-size dependence of the initial relaxation time steps and autocorrelation time steps.
    The \textcolor{black}{dark cyan} empty [filled] circles represent the initial relaxation time steps for $(m, n, k) = (1, 9, 1)$ [$(m, n, k) = (1, 7, 0)$].
    The brown empty [filled] diamonds represent the autocorrelation time steps for $(m, n, k) = (10, 11, 1)$ [$(m, n, k) = (1, 8, 0)$].
    }
\label{fig:relaxation_autocorrelation}
\end{figure}

In this appendix, we discuss how the initial relaxation and autocorrelation time steps depend on system size (see Secs.~\ref{subsec:initical-relaxation} and \ref{subsec:autocorrelation} for the definitions of these quantities). Figure~\ref{fig:relaxation_autocorrelation} presents numerical results for various combinations of $(m, n, k)$ values, with the model and conditions consistent with those in Sec.~\ref{sec:numerical-demonstration} except for system-size variations.
In this figure, the \textcolor{black}{dark cyan} empty [filled] circles represent the initial relaxation time steps for $(m, n, k) = (1, 9, 1)$ [$(m, n, k) = (1, 7, 0)$], while the brown empty [filled] diamonds represent the autocorrelation time steps for $(m, n, k) = (10, 11, 1)$ [$(m, n, k) = (1, 8, 0)$].
These results indicate that both the initial relaxation and the autocorrelation time steps do not scale exponentially with system size. This finding suggests that our MCMC-SPU algorithm is scalable, at least within the scope of our simulations.

\section{Detailed data of the successful postselection probability}\label{apppend:detail_PS_prob}

\begin{figure*}[htbp]
\includegraphics[keepaspectratio, scale=1.37]{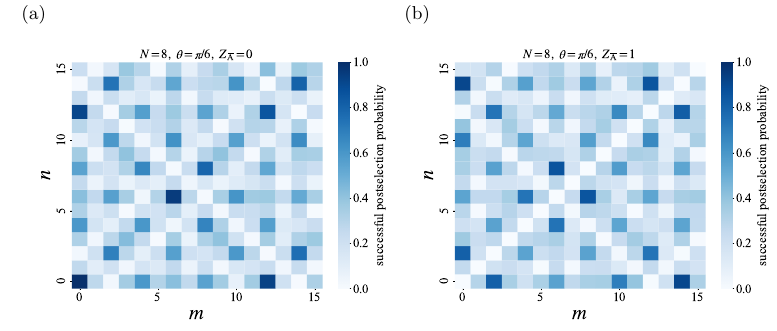}
\caption{
    Successful postselection probability for each quantum circuit in the MCMC-SPU algorithm for the one-dimensional transverse-field Ising model with eight sites.
    The plots show the probability of successful postselection of measurement outcomes on the ancilla qubits conditioned on the measurement outcome of the ancilla qubit for superposition with index $\overline{\rm A}$ being (a) $Z_{\overline{\rm A}}=0$  and  (b) $Z_{\overline{\rm A}}=1$, respectively, for quantum circuit $U_{m n}$. This quantum circuit incorporates unitary gates $U_m$ and $U_n$ corresponding to a pair of orders $m$ and $n$ stochastically sampled from the expansion of an imaginary-time evolution.
    In Fig.~(b), although the results are plotted as zero for $m=n$, this representation is solely for convenience. It actually indicates that no MCMC sampling is conducted for these specific index combinations (see Eq.~\eqref{eq:W_no_contribution} and the following argument).
    }
    \label{fig:sampling_PSprob}
\end{figure*}

We provide detailed data on the successful postselection probability for each quantum circuit in the MCMC-SPU algorithm, which is discussed in Sec.~\ref{subsec:probability-PS}. 
Our numerical simulation focuses on the one-dimensional transverse-field Ising model specified in Eq.~\eqref{eq:Hamiltonian-QTFI}  with $N=8$ and $\theta = \pi/6$ as an illustrative example.
Figure~\ref{fig:sampling_PSprob} illustrates the numerical results of the successful postselection probability for each quantum circuit $U_{m n}$ in the MCMC-SPU algorithm for the transverse-field Ising model. Figure~\ref{fig:sampling_PSprob}(a) [(b)] shows the probability of successful postselection of
measurement outcomes on the ancilla qubits, which is evaluated conditioned on the measurement outcome of the ancilla qubit for superposition with index $\overline{\rm A}$ being $Z_{\overline{\rm A}}=0$ [$Z_{\overline{\rm A}}= 1$] for the quantum circuit $U_{m n}$. This quantum circuit incorporates unitary gates $U_m$ and $U_n$ corresponding to a pair of orders $m$ and $n$ stochastically sampled from the expansion of an imaginary-time evolution (Steps~1--2 in Sec.~\ref{subsec:MCMC-SPU-procedure}).
These plots confirm that appropriate postselection is achieved with nonzero probability across most combinations of indices $(m, n, Z_{\overline{\rm A}})$ except for the cases $(m, m, 1)$.
Here, although the results are plotted as zero for these cases, this representation is solely for convenience. It actually indicates that no MCMC sampling is conducted for these specific index combinations (see Eq.~\eqref{eq:W_no_contribution} and the following argument).
The unweighted average of the results shown in Fig.~\ref{fig:sampling_PSprob}  yields the results in Fig.~\ref{fig:resource_implementation}(d) in Sec.~\ref{subsec:probability-PS}.

\section{Analysis of the initial relaxation}\label{append:Gelman-Rubin}

In this appendix, we delve into the technical details of analyzing the initial relaxation time step in our MCMC-SPU algorithm, as discussed in Sec.~\ref{subsec:initical-relaxation}. Our analysis leverages the Gelman-Rubin statistics~\cite{Gelman1992, Gelman2003, Vats2021}, which we briefly review here.
Generally, we consider a set of data series $\{\theta_{i j}\}_{i j}$, where $\theta_{i j}$ denotes the $i$-th element of the series labeled with $j$.
In our specific analysis in Sec.~\ref{subsec:initical-relaxation}, there are two series: label $j = 1$ corresponds to the initial state $\ket{\psi_{\rm F}}$ and $j=2$ corresponds to $\ket{\psi_{\rm AF}}$ for the Markov process, while label $i$ represents the time step in each series. We generate $2N_{\rm step}$ time steps, discarding the first $N_{\rm step}$ steps as a burn-in period. We then consider the statistical ensemble consisting of the results of the last $N_{\rm step}$ steps in the following.

We first define the within-chain variance as follows:
\begin{align} \label{eq:W_GR}
W_{\rm GR} := \frac{1}{N_{\rm step}(N_{\rm series}-1)} \sum_{j=1}^{N_{\rm series} } \sum_{i=1}^{N_{\rm step} } \qty(\theta_{i j} - \overline{\theta}_j)^2,
\end{align}
where $\overline{\theta}_j$ denotes the mean value of series $\{\theta_{i j}\}_i$, i.e.,
\begin{align}
\overline{\theta}_j
:= \frac{1}{N_{\rm step}} \sum_{i = 1}^{N_{\rm step}} \theta_{i j}.
\end{align}
We also define the between-chain variance as follows:
\begin{align} \label{eq:B_GR}
B_{\rm GR}
:= \frac{N_{\rm step}}{N_{\rm series}-1} \sum_{j=1}^{N_{\rm series}}  \qty(\overline{\theta}_j - \overline{\theta})^2,
\end{align}
where $\overline{\theta}$ denotes the mean value of all data in the $N_{\rm series}$ series, i.e.,
\begin{align}
\overline{\theta}
:= \frac{1}{N_{\rm step} N_{\rm series}}\sum_{j=1}^{N_{\rm series}} \sum_{i=1}^{N_{\rm step}} \theta_{i j}
= \frac{1}{ N_{\rm series}}\sum_{j=1}^{N_{\rm series}}\overline{\theta}_{ j}.
\end{align}

Building on the two types of variance defined in Eqs.~\eqref{eq:W_GR} and \eqref{eq:B_GR}, we calculate the estimator $\hat{V}_{\rm GR}$ for the variance of the stationary distribution consisting of the whole data from all the series as follows:
\begin{align}
\hat{V}_{\rm GR} (N_{\rm step})
= \frac{N_{\rm step}-1}{N_{\rm step}} W_{\rm GR} + \frac{1}{N_{\rm step}} B_{\rm GR}.
\end{align}
We then define the potential scale reduction factor $\hat{R}_{\rm GR} (N_{\rm step})$ as follows:
\begin{align}
\hat{R}_{\rm GR} (N_{\rm step})
:= \sqrt{\frac{\hat{V}_{\rm GR} (N_{\rm step})}{W_{\rm GR}}}.
\end{align}
If $\hat{R}_{\rm GR} (N_{\rm step}) \le 1.1$, it indicates that $N_{\rm series}$ series are statistically indistinguishable, confirming convergence has been achieved~\cite{Gelman2003, Vats2021}.
In our specific analysis in Sec.~\ref{subsec:initical-relaxation},
we conclude that $N_{\rm step}$ is sufficient as the burn-in period if this condition is satisfied.\\

\section{Jackknife method and autocorrelation}\label{appned:jackknife-autocorrelation}

In this appendix, we present a brief review of the general theory of the jackknife method~\cite{Quenouille1949_jackknife, Quenouille1956, Tukey1958, Efron1994}, as used in the analysis of the autocorrelation time step in our MCMC-SPU algorithm in Sec.~\ref{subsec:autocorrelation}.
The jackknife method, applied with a bin size $s_{\rm bin}$, helps evaluate autocorrelation in a data sequence.
We first divide the series $\{A_l\}_l$ of $M_{\rm data}$ data into $M_b = M_{\rm data} /s_{\rm bin}$ bins and then define $\ev{A}_b$ for each label $b$ of bin as
\begin{align}
\ev{A}_b
:= \frac{1}{M_{\rm data} - s_{\rm bin}} \sum_{l \notin B_b} A_l,
\end{align}
where $B_b$ denotes the bin of data labeled $b$.
We then define the mean of the function $f(A)$ of the data $A$ as
\begin{align}
\ev{f(A)}
&= \frac{1}{M_b} \sum_{b=1}^{M_b} f(\ev{A}_b)
\end{align}
with its standard error
\begin{align}
\delta\ev{f(A)}
&= \sqrt{(M_b-1) \qty[\ev{f(A)^2} - \ev{f(A)}^2]}.
\end{align}
When the bin size $s_{\rm bin}$ is sufficiently large compared with the autocorrelation time step $\tau$, bins with different labels can be considered uncorrelated, and the standard error $\delta\ev{f(A)}$ becomes independent of the bin size $s_{\rm bin}$.
On the basis of this observation, we gradually increase $s_{\rm bin}$ and examine the $s_{\rm bin}$-dependence of $\delta\ev{f(A)}$.
If the value of $\delta\ev{f(A)}$ saturates at some value $s_{\rm bin}$ and remains consistent within fluctuations, the integrated autocorrelation time step is estimated as $\tau = s_{\rm bin}/2$.
In our specific analysis in Sec.~\ref{subsec:autocorrelation}, data $A$ corresponds to the label of the state of the target system in the computational basis, and function $f(A)$ corresponds to the value of an observable.

\section{\textcolor{black}{Classical computational overhead}}\label{append:classical-overhead}

\begin{table*}
	\centering
	\textcolor{black}{\caption{
    	Summary of the classical computational overhead and its scaling.}
	\label{table:classical-overhead-scaling}
    }
	\vspace{1mm}
    \textcolor{black}{
	\begin{tabular}{|c||c|c|c|}
	\hline
    	& (a) \textcolor{black}{ quantum-circuit sampling} &  (b) post-processing &
	(c) monitoring convergence
	  \\
	\hline\hline
    \begin{tabular}{l}
	{quantity characterizing}\\
	{the overhead}
	\end{tabular}
 &  
 number $N_{\epsilon, \delta}$
of sampling
& \begin{tabular}{c}
	{number of samples}\\
	{to evaluate $Z_{m n}^k$}
	\end{tabular}
 &  \begin{tabular}{c}
	{initial relaxation 
and autocorrelation}\\
	{time steps \textcolor{black}{for each quantum circuit}}
	\end{tabular}
\\
	\hline
	temperature dependence & 
	$N_{\epsilon, \delta}\propto e^{2\beta}$
	  & No & No \\
	\hline
	system-size dependence & 
	No \ [see Eq.~\eqref{eq:number_Qcircuit}]
	 & \begin{tabular}{c}
	{not significantly}\\
	{increase with size}
	\end{tabular} & \begin{tabular}{c}
	{not significantly}\\
	{increase with size}
	\end{tabular} \\
	\hline
	scalability & 
	 Yes 
	& Yes \ [see Appendix~\ref{append:reweighting}] & Yes \ [see Appendix~\ref{append:scaling_overhead}] \\
	\hline
	\begin{tabular}{l}
	{trade-off with}\\
	{quantum resource}\\
    {\textcolor{black}{(compared to}}\\
    {\textcolor{black}{conventional QMETTS)}}
	\end{tabular} &
\textcolor{black}{
\begin{tabular}{l}
	additional overhead of $N_{\epsilon, \delta}$\\
    while reducing\\the average circuit depth	\end{tabular}
    }
	& 
	\begin{tabular}{l}
	additional overhead\\
	of evaluating $Z_{m n}^k$\\
    while reducing\\
    the circuit-execution time
	\end{tabular}
	 & \begin{tabular}{l}
	a few-times increase\\
	in the above time steps\\
    while reducing\\
    the circuit-execution time
	\end{tabular}\\
	\hline
	\end{tabular}
    }
\end{table*}

\textcolor{black}{In this appendix, we discuss the classical computational overhead associated with our MCMC-SPU algorithm and its scaling. We also consider a potential strategy for mitigating the overhead.}

\textcolor{black}{\subsection{Scaling of the overhead}
}

\textcolor{black}{\textcolor{black}{In Table~\ref{table:classical-overhead-scaling}, we summarize the classical computational overhead associated with our algorithm and its scalability, i.e., avoiding exponential overhead growth with increasing system size.} We discuss \textcolor{black}{these aspects} in the following, particularly focusing on (a) \textcolor{black}{quantum-circuit sampling}, (b) post-processing, and (c) monitoring convergence.}

\textcolor{black}{(a) First, we consider \textcolor{black}{quantum-circuit sampling}, equivalent to sampling pairs of orders of polynomials (see Steps~1.--2. in Sec.~\ref{subsec:MCMC-SPU-procedure}). This overhead is characterized by the number $N_{\epsilon, \delta}$ of sampling. \textcolor{black}{As mentioned below Eq.~\eqref{eq:number_Qcircuit}, this quantity does not include sampling overhead associated with calculating the quantum mechanical expectation value.} As described in Eq.~\eqref{eq:number_Qcircuit}, this number scales as $N_{\epsilon, \delta}\propto e^{2\beta}$ with inverse temperature $\beta$. The number $N_{\epsilon,\delta}$ shows no explicit dependence on the system size, ensuring scalability. \textcolor{black}{In exchange for this additional sampling overhead, the scaling of the average circuit depth $d_{\rm average}\propto\sqrt{\beta}$ is improved from $d\propto\beta$ in the conventional QMETTS algorithm [see Eqs.~\eqref{eq:required-order-LCU}--\eqref{eq:average-order-LCU}].}
}

\textcolor{black}{(b) Next, we consider the post-processing. This overhead is characterized by the number of samples required to evaluate the weight factor $Z_{mn}^k$, defined in Eq.~\eqref{eq:MCMC-SPU-total-weight}. As detailed below Eq.~\eqref{eq:MCMC-SPU-each-ensemble-average}, this quantity can be obtained as a statistical mean of uniformly sampled quantities $\{W_{mn}^{ik}\}$, representing the probabilities of successful postselection of the corresponding measurement outcomes of the ancilla qubit [see Eq.~\eqref{eq:MCMC-SPU-PSprob}]. This uniform sampling is independent of the temperature since the temperature dependence only appears in the coefficients $\{c_n\}$. The system-size dependence of the number of samples associated with this evaluation is analyzed in Appendix~\ref{append:reweighting}. \textcolor{black}{Specifically, Fig.~\ref{fig:stdev_weight} shows} that the distribution of the values of ${W_{mn}^{ik}}$ remains nearly uniform with increasing system size, indicating scalability, \textcolor{black}{i.e., avoiding exponential overhead growth. While incurring this additional overhead unique to our algorithm, the execution time for each quantum circuit is reduced compared to the conventional QMETTS algorithm.}}

\textcolor{black}{(c) Then, we consider the monitoring convergence. This overhead is characterized by the initial relaxation and autocorrelation time steps, introduced in Secs.~\ref{subsec:initical-relaxation} and \ref{subsec:autocorrelation}. These time steps \textcolor{black}{for each quantum circuit} are independent of the temperature since the temperature dependence only appears in the coefficients $\{c_n\}$. \textcolor{black}{Nevertheless, near the critical temperature, there is a possibility that several kinds of quantum circuits have significant contributions, potentially relevant to nontrivial scaling behaviors associated with critical phenomena. A detailed investigation in this direction is beyond the scope of this paper and is left for an interesting future work.} The system-size \textcolor{black}{dependences of these time steps} are examined in Appendix~\ref{append:scaling_overhead}. \textcolor{black}{Specifically, Fig.~\ref{fig:relaxation_autocorrelation} shows} that these quantities do not \textcolor{black}{exponentially} increase with system size, indicating scalability. In exchange for a few-times increase in these time steps compared with the conventional QMETTS algorithm, a reduction in the execution time for each quantum circuit is achieved.}

\textcolor{black}{In summary, all aspects (a)--(c) discussed above are scalable and exhibit trade-off relations with quantum computational resources.}

\textcolor{black}{\subsection{Potential strategy for mitigating the overhead}}

\textcolor{black}{We discuss a potential strategy for mitigating the classical computational overhead. In general, as with the examples (a)--(c) above, the classical computational overhead involves evaluating the statistical mean of sampled data. The scaling of the sampling complexity associated with such mean evaluation can be improved by employing, for example, the median-of-means (MoM) technique~\cite{Nemirovskii1983, Birge1984, Lerasle2019}. In this technique, instead of performing uniform sampling, the total number $n_{\rm tot}$ of samples is divided into $n_{\rm block}$ blocks of size $n_{\rm size}$, i.e., $n_{\rm tot}=n_{\rm block}\ n_{\rm size}$. The MoM estimator ${\hat{\mu}}_{\rm MoM}$ is defined as the median of a series of the statistical means ${\hat{\mu}}_l$ of each block, i.e.,
\begin{align}
{\hat{\mu}}_{\rm MoM} := {\rm median} ({\hat{\mu}}_1, {\hat{\mu}}_2, \cdots, {\hat{\mu}}_{n_{\rm block}}).
\end{align}
The probability that this estimator deviates from the true mean $\mu_0$ by more than $\varepsilon$ is shown to be bounded as~\cite{Nemirovskii1983, Birge1984, Lerasle2019}
\begin{align}
{\rm Prob}\qty(|{\hat{\mu}}_{\rm MoM} - \mu_0|>\varepsilon)\le \exp\!\qty(\! -2 n_{\rm block} \qty[\frac{1}{2} - \frac{\sigma^2}{n_{\rm size} \varepsilon^2}]^2),
\end{align}
where $\sigma^2$ denotes the variance of sampled data. This inequality shows that the following sampling complexity suffices for successful estimation with probability at least $1 - \nu$: 
\begin{align}
n_{\rm block}&=\mathcal{O}\qty(\log \qty[\nu^{-1}] ), \\ n_{\rm size}&=\mathcal{O}\qty(\varepsilon^{-2}).
\end{align}
Thus, the total sampling complexity of the MoM technique
\begin{align}
n_{\rm tot}=n_{\rm block} n_{\rm size} = \mathcal{O}\qty( \varepsilon^{-2} \log \qty[\nu^{-1}] )
\end{align}
is exponentially improved in $\nu^{-1}$ compared with the counterpart
$
\mathcal{O} \qty( \varepsilon^{-2} \nu^{-1} )
$
required for a naive uniform sampling. The latter is obtained by the Chebyshev’s inequality, applicable to general quantities, including ones outside the scope of the Hoeffding's inequality~\cite{Hoeffding1963}.
}

\bibliography{MCMC-SPU_v2}

\end{document}